\documentclass[a4paper,11pt]{article}
\pdfoutput=1 
\usepackage{jcappub} 
\usepackage[T1]{fontenc} 
\usepackage{graphicx}   
\usepackage{changepage}
\usepackage{subcaption}
\usepackage{orcidlink}
\usepackage{subcaption}
\usepackage{academicons}
\usepackage{natbib}
\usepackage{hyperref}
\usepackage{aas_macros}
\definecolor{orcidlogocol}{HTML}{A6CE39}

\newcommand{\Change}[1]{{\color{blue}[Change: #1]}}

\title{Turbulence Induced Non-Gaussian Spectral Distortion in the Microwave Sky from Photon-Axion Conversion in Galaxy Clusters}

\author{Harsh Mehta\orcidlink{0009-0007-4664-4820},}
\author{and Suvodip Mukherjee\orcidlink{0000-0002-3373-5236}}

\affiliation{Department of Astronomy and Astrophysics, Tata Institute of Fundamental Research, Homi Bhabha Road, Mumbai, 400005, India}

\emailAdd{harsh.mehta@tifr.res.in}
\emailAdd{suvodip@tifr.res.in}

\begin{document}

\abstract{The conversion of CMB photons to axions (or axion-like particles (ALPs)) can lead to a unique spectral distortion in the temperature and polarization sky which can be explored in upcoming CMB experiments. In this work we have developed a numerical simulation-based technique of photons to ALPs conversion in the galaxy clusters and show for the first time that this physical process can lead to large non-Gaussian signal in the temperature and polarization field, which is impacted by the presence of inhomogeneities and turbulence in the electron density and magnetic field. Our simulation-based technique can simulate the theoretical signal for different scenarios of cluster electron density and magnetic field turbulence and provides testable predictions to discover ALPs from galaxy clusters using spatially non-Gaussian and anisotropic spectral distortion of the microwave sky. We show that the presence of turbulence in the magnetic field and electron density can impact the Gaussian part of the signal captured in terms of the angular power spectra of the signal by more than an order of magnitude. Also, the presence of turbulence in different clusters will lead the temperature and polarization fluctuations around the cluster region to have varying non-Gaussian distribution, with peaks and tails different from the Gaussian statistics of the CMB anisotropy. This new numerical technique has made it possible to calculate also the non-Gaussian signals and can be used in future CMB analysis in synergy with X-ray and radio observations to unveil ALPs coupling with photons in the currently unexplored ranges, for the masses between about $10^{-14}$ eV--$10^{-11}$ eV.}

\maketitle
\section{Introduction}
The Cosmic Microwave Background (CMB) follows an almost ideal black-body spectrum with a temperature of $\sim 2.7255$ K \cite{Dodelson:2003ft,Fixsen_1996,fixsen1996cosmic}. It bears deviations in temperature as well as polarization at different sky locations, of low order ($\sim 10^{-5}$) \cite{Dodelson:2003ft,Fixsen_1996,fixsen1996cosmic,hanson2009estimators}. Also, there are spectral distortions in the CMB that refer to the tiny deviations from the ideal black-body spectrum \cite{2014PTEP.2014fB107T,erler2018planck,chluba2012evolution,lucca2020synergy}. The CMB spatial anisotropies and spectral distortions contain information on a variety of signals \cite{Dodelson:2003ft,hu2002cosmic,lucca2020synergy,chluba2012evolution,erler2018planck,2014PTEP.2014fB107T,hu1995toward,hu1997cmb,hu1998complete,lewis2004cmb}, with their distinguishing variations in the spectral and spatial domains, such as the Sunyaev-Zeldovich (SZ) effects \cite{1972CoASP...4..173S}, lensing \cite{Smith_2007,lewis2006weak}, reionization \cite{adam2016planck,barkana2001beginning}, etc., within the standard model of cosmology and particle physics. The distortions in CMB spatial and spectral properties can also arise from Beyond Standard Model (BSM) physics, making it possible to discover uncharted territories \cite{Cyr:2024sbd,chluba2016spectral,chluba2021new,shoemaker2016probing}.

One such BSM signature arises from axions or axion-like particles (ALPs) which are hypothetical particles predicted by various theories that can be dark matter candidates as they possess a small mass and have a weak interaction with photons \cite{dine1983not,abbott1983cosmological,preskill1983cosmology,Ghosh:2022rta,1992SvJNP..55.1063B,khlopov1999nonlinear,sakharov1994nonhomogeneity,sakharov1996large}. They can be probed using the weak coupling they may be having with photons in the presence of magnetic field. Such signals can arise from astrophysical objects with magnetic fields such as galaxies, galaxy clusters, and voids due to the  conversion of CMB photons to ALPs, if ALPs exist in nature, irrespective of whether they constitute dark matter or not. The resonant conversion of photons to ALPs leads to a polarized spectral distortion (departure from Planck black-body spectrum), referred to as the $\alpha-$distortion in the CMB, which can be probed using CMB observations of the temperature and polarization fluctuations \cite{Mukherjee_2020,osti_22525054} in the form of a secondary anisotropy. 
This will result in a change in the shape of the CMB black-body spectrum as well as the power spectra of spatial anisotropies in the temperature and polarization (E-mode and B-mode) power spectra \cite{Mehta:2024pdz,Mehta:2024wfo,mondino2024axioninducedpatchyscreeningcosmic,Goldstein:2024mfp}. 

The ALP distortion signal from a galaxy cluster depends on its electron density and magnetic field profiles, deciding not only its strength but also the shape and spatial extent of this distortion signal from the cluster region.
In our earlier works \cite{Mukherjee_2020,Mehta:2024wfo,Mehta:2024sye,Mehta:2024pdz}, we have shown how the ALP signal is impacted in the case of smooth radially varying coherent magnetic fields and electron densities in galaxy clusters. In general, the variation in the profiles will not be smooth and can have inhomogeneities associated with their magnitudes. Also, the magnetic fields may not always be coherent on large scales. This occurs as a result of the turbulence caused by the energy injection as a result of the various gravitational and astrophysical phenomena that affect the underlying density or velocity fields in galaxy clusters \cite{subramanian2006evolving,schekochihin2006turbulence,xu2009turbulence,ensslin2006magnetic,xu2012comparisons}. 

In this work, we demonstrate for the first time the effects of these inhomogeneities and incoherence in the electron density and magnetic field on the ALP signal by using a three-dimensional simulation setup that models these effects from length scale of a few parsecs to mega-parsecs. The simulation setup can simulate the ALP signal for different ALP masses in the range $10^{-14} - 10^{-11}$ eV for different coupling strengths \cite{Mehta:2024pdz,Mehta:2024wfo}. This setup will enable us to theoretically calculate the ALP distortion signal for both temperature and polarization anisotropy in CMB for any generic turbulent electron density and magnetic field. {We analyze the effects of turbulence in electron density and magnetic field in galaxy clusters on the ALP signal using different scenarios of incoherence or strength inhomogeneity. We discuss the distinct signature of ALPs in the form of the spectral variation of temperature-polarization (TP) cross spectrum, which can be robustly used, without any impact from spatial turbulent effects. Also, we analyze the non-Gaussianity effect of turbulence on the ALP signal, which can be used to distinguish it from the CMB \cite{aghanim1999searching,hanson2009estimators,challinor2012cmb,liguori2006testing,ade2016planck}. }
This setup can take into account realistic cluster scenarios and even extreme ones, characterized by very high turbulence in not-so-relaxed systems \cite{subramanian2006evolving,schekochihin2006turbulence,xu2009turbulence,ensslin2006magnetic,xu2012comparisons}. The simulation setup can also use the magnetic field and electron density profiles from large-scale hydrodynamical simulations, such as IllustrisTNG, GADGET, CosmoMHD, etc. \cite{Marinacci_2018,dolag2009mhd,li2008cosmomhd}, to calculate the ALP signal from galaxy clusters on a cosmological scale. It can be integrated with our previously developed code \texttt{SpectrAx} for data analysis using multi-band observations \cite{Mehta:2024sye}, to take into account the impact of inhomogeneities and incoherence on the ALP distortion signal and mitigate its impact in inferring the photon-ALP coupling constant $g_{a\gamma}$ from observations. The simulation setup can be applied not only to galaxy clusters, but also to analyze data from other astrophysical systems such as neutron stars and galactic halos.

The paper is organized as follows: the photon-ALP resonant conversion is explained in Sec. \ref{sec: alp signal}. The effect of turbulence in cluster profiles is discussed in Sec. \ref{sec:inhomogeneity}. This is followed by a description of the simulation setup in Sec. \ref{sec:method}. The ALP signal in temperature and polarization is dealt with in Sec. \ref{sec:powspectra}, and the results for different inhomogeneity cases are shown in Sec. \ref{sec:results}. The pros of this simulation setup are highlighted in Sec. \ref{sec:implications}, followed by a summary of the work in Sec. \ref{sec:conclude}. This work mostly uses natural units ($\hbar = 1, c = 1, k_B = 1$), unless explicitly stated. For the cosmological parameters, we have used the values from Planck 2018 \cite{aghanim2020planck}.

\section{The ALP signal in the CMB}
\label{sec: alp signal}
If ALPs exist, the CMB photons travelling through galaxy clusters may get converted to ALPs in the transverse magnetic field of the ionized plasma in the intra-cluster medium (ICM), depending on its coupling strength denoted by $g_{a \gamma}$. This conversion may be resonant or non-resonant in nature, with the resonant ones being stronger and needing the mass resonant condition to be satisfied, i.e.,
\begin{equation}
 m_a = m_{\gamma} = \frac{\hbar \omega_p}{c^2} \approx \frac{\hbar}{c^2}\sqrt{n_e e^2 / m_e \epsilon_{0}}, 
\label{eq:resonance mass}
\end{equation}
with $\omega_p$ being the plasma frequency, and $n_e$ the electron density at the conversion location \cite{Mukherjee_2020}. 
ALPs of masses $10^{-14}$ to $10^{-11}$ eV can be probed based on the observed electron densities in galaxy clusters. The Lagrangian for the photon-ALP interaction is given as \cite{Raffelt:1996wa}:
\begin{equation}\label{eq:lagr}
  \mathcal{L}_{\mathrm{int}} = -\frac{g_{a \gamma} F_{\mu \nu}\tilde{F}^{\mu \nu} a}{4} = g_{a \gamma} \overrightarrow{E} \cdot \overrightarrow{B}_{\mathrm{ext}} a,
\end{equation}
with $F_{\mu \nu}$ being the electromagnetic field tensor, $\tilde{F}^{\mu \nu}$ the dual tensor, and $a$ the ALP field. The Lagrangian suggests the involvement of the transverse magnetic field in the conversion and no conversion in the presence of a longitudinal field. 
We define the parameters for the variation of electron density ($\Delta_e$) and conversion scale of photon-ALP system ($l_{\mathrm{osc}} = 2\pi / \Delta_{\mathrm{osc}}$) as:
\begin{equation}
\label{eq:lenparams}
\begin{split}
\Delta_a &= - m_a^2 / 2\omega  \,,
\qquad
\Delta_e \approx - \omega_p^2 / 2\omega \,,
\\
\Delta_{a\gamma} &= g_{a\gamma}B_{t} / 2 \,,
\qquad
\Delta_{\mathrm{osc}}^2 = (\Delta_a - \Delta_e)^2 + 4\Delta_{a\gamma}^2 .
\end{split}
\end{equation}

The probability of conversion is related to the adiabaticity parameter ($\gamma_{\mathrm{ad}}$), which compares the length scale of the photon-ALP oscillation to that of the variation of electron density. In the adiabatic limit ($\gamma_{\mathrm{ad}} >> 1$), the conversion will be complete, whereas in the non-adiabatic limit ($\gamma_{\mathrm{ad}} << 1$), the conversion scale is greater than the scale of variation of electron density, and the Landau-Zener approximation gives:
\cite{Mukherjee_2020,Mehta:2024wfo,osti_22525054}
\begin{equation}
P_{\mathrm{conv}} = 1 - e^{-\pi \gamma_{\mathrm{ad}}/2} \approx \pi \gamma_{\mathrm{ad}} / 2, 
\label{eq:prob}
\end{equation} 
with the adiabaticity parameter being:
\begin{equation}
\gamma_{\mathrm{ad}} = \frac{\Delta_{\mathrm{osc}}^2}{|\nabla \Delta_{e}|} = \left| \frac{2g_{a\gamma }^2 B_t^2 \nu (1 + z)}{\nabla \omega_p^2} \right|.
\label{eq:gamma_ad}
\end{equation}
Here, $B_t$ is the transverse magnetic field at the resonant location, and $\nu(1 + z)$ is the frequency of photons at the cluster redshift. The effect of Faraday rotation is neglected as it will be negligible in the microwave spectrum. 

For a spherically symmetric cluster with homogeneous electron density and magnetic field profiles, the ALPs of a particular mass are formed within a spherical shell in the cluster. This shell is observed as a signal disk when projected on the two-dimensional sky. The signal generally increases in the outer regions of the disk and then decays to zero based on the extent of the disk. For such a cluster, there will only be two resonances for a particular mass ALP along the line of sight, and the ALP signal will have the probability corresponding to only one conversion \cite{Mukherjee_2020}. 
However the assumption of homogeneous and coherent profiles is too idealized, which will be affected by the astrophysical and gravitational phenomena in clusters. This will lead to strength inhomogeneity and incoherence in the electron density and magnetic field profile in galaxy clusters.

The effects of turbulence in cluster profiles will impact the spatial features of the ALP signal in terms of its shape, spatial extent, and strength. 
If the electron density profile is not homogeneous, there may be multiple resonances for a particular mass ALP. This will distort the shape of the disk, while also changing its spatial extent. There will be a conversion probability related to each resonance ($P_{\mathrm{conv}}$), but the effective conversion probability through the cluster will be the sum of probabilities of the number of odd conversions possible.  
For photons traveling through a cluster, the net conversion probability to leading order, along a line of sight, can then be given as:
\begin{equation}
    P(\gamma \rightarrow a) \approx \frac{\pi}{2} \sum_{i = 1}^{N}  \gamma_{\mathrm{ad},i} \, ,
   \label{eq: Probab}
\end{equation}
with the adiabaticity parameters being calculated at the resonant locations which are $N$ in number. The effect of turbulence in electron density will also affect the strength of the ALP signal, changing the resonant locations as well as the conversion probability at all those locations due to the gradient of the plasma frequency ($\gamma_{\mathrm{ad}} \propto 1 / 
|\nabla n_e|$). 
 Strength inhomogeneity in magnetic field  will impact the strength of the signal at the resonant locations by changing the conversion probability ($\gamma_{\mathrm{ad}} \propto 1 / B_t^2$). 
 There will also be accompanying polarization signals from multiple resonances, that will depend on the transverse magnetic field directions at those locations.
Incoherence in the magnetic field profile will lead to depolarization of the ALP signal as the polarized contributions from various resonant locations will superpose. Hence, the electron density variation scale and the coherence length of the magnetic field play a crucial role in the spatial features of the ALP signal.
Thus, we need to account for the effects of turbulence in cluster profiles when estimating the ALP signal, as otherwise it can lead to biases in the constraining of the ALP coupling constant $g_{a\gamma}$ \cite{2017,Mirizzi_2009,Mukherjee_2019,Mehta:2024sye,Mehta:2024wfo,Mehta:2024pdz,mondino2024axioninducedpatchyscreeningcosmic,Goldstein:2024mfp}. 

\section{Modelling the turbulence in electron density and magnetic field}
\label{sec:inhomogeneity}
The properties of galaxy clusters 
are driven by the various astrophysical and gravitational phenomena taking place in them \cite{subramanian2006evolving,schekochihin2006turbulence,xu2009turbulence,ensslin2006magnetic,donnert2018magnetic}. These processes inject energy and affect the hydrodynamics of the ionized plasma in the ICM, and also the cores of clusters.
The energy injection into the ICM is mainly expected to take place via cluster mergers. This affects the density and velocity fields of clusters, which results in a change in the hydrodynamics of the associated ionized plasma. If the Reynoldes number (ratio of the inertial to viscous forces) becomes high, there will be turbulence that will lead to the cascading of this energy from large scales (low wave numbers $k$) to small scales (high $k$ values), which will then be lost due to viscous dissipation. 
The presence of turbulence leads to the dynamo effects in clusters, which amplify the cluster magnetic fields. This magnetic field then affects the plasma hydrodynamics, hence the two are inter-linked \cite{GOVONI_2004,condon2016essential,carilli2002cluster,clarke2001new,Ferrari_2008}. 
 This complexity related to the plasma hydrodynamics leads to clusters having different profiles that determine their electron densities, magnetic fields, temperatures, etc. It is these profiles that determine the various emissions taking place from the ICM, such as the synchrotron emission, SZ effects, etc. These profiles, thus, cannot always be approximated to be homogeneous with a smooth radial variation when modelling the various signals from a cluster. A turbulent component will always affect these smooth profiles and hence affect the various emissions that take place from clusters. There will also be an associated incoherence in the direction of magnetic fields in clusters, which may lead to a direction-based effect on the polarization or the intensity of photons with respect to the line of sight. This turbulence affects the cool cores of galaxy clusters as well, which need a mechanical dissipation of energy from radio bubbles to maintain their low temperatures.
 
 Various hydrodynamical simulations, such as IllustrisTNG, GADGET, CosmoMHD, etc., account for these effects on the cluster profiles, but these simulations typically probe kpc scale and do not resolve subpc scale turbulence \cite{Marinacci_2018,dolag2009mhd,li2008cosmomhd,donnert2018magnetic,ruszkowski2011cosmological,xu2012comparisons}. A more concrete basis for the study of the effects of inhomogeneities in cluster profiles would require higher resolutions of the order of subpc and models that take into account the effects of high turbulence at small scales in the non-linear regime \cite{subramanian2006evolving,schekochihin2006turbulence,xu2009turbulence,ensslin2006magnetic}.

 \begin{figure}[h!]
     \centering
\includegraphics[height=7cm,width=11cm]{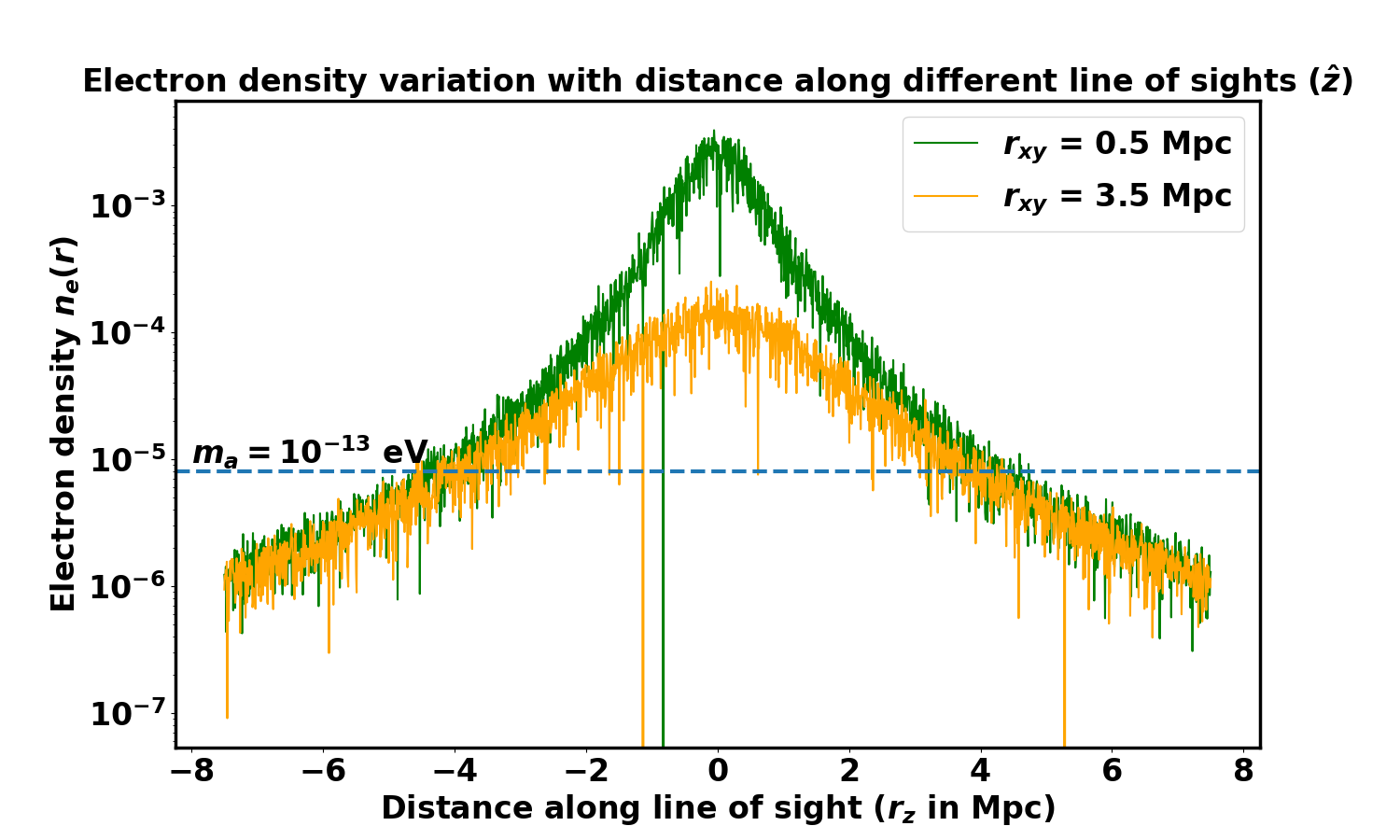}
    \caption{Electron density magnitude (with turbulence standard deviation being 30\% of the mean radial profile) for different locations along the line of sight for two different projected distances. The horizontal dashed line corresponds to the electron density required for the resonant condition to be satisfied for $10^{-13}$ eV mass ALPs, which happens at different longitudinal locations for the two cases shown. The length of the setup considered here is 15 Mpc.}
       \label{fig:ne_los}
\end{figure}

\begin{figure}[h!]
     \centering
\includegraphics[height=7cm,width=11cm]{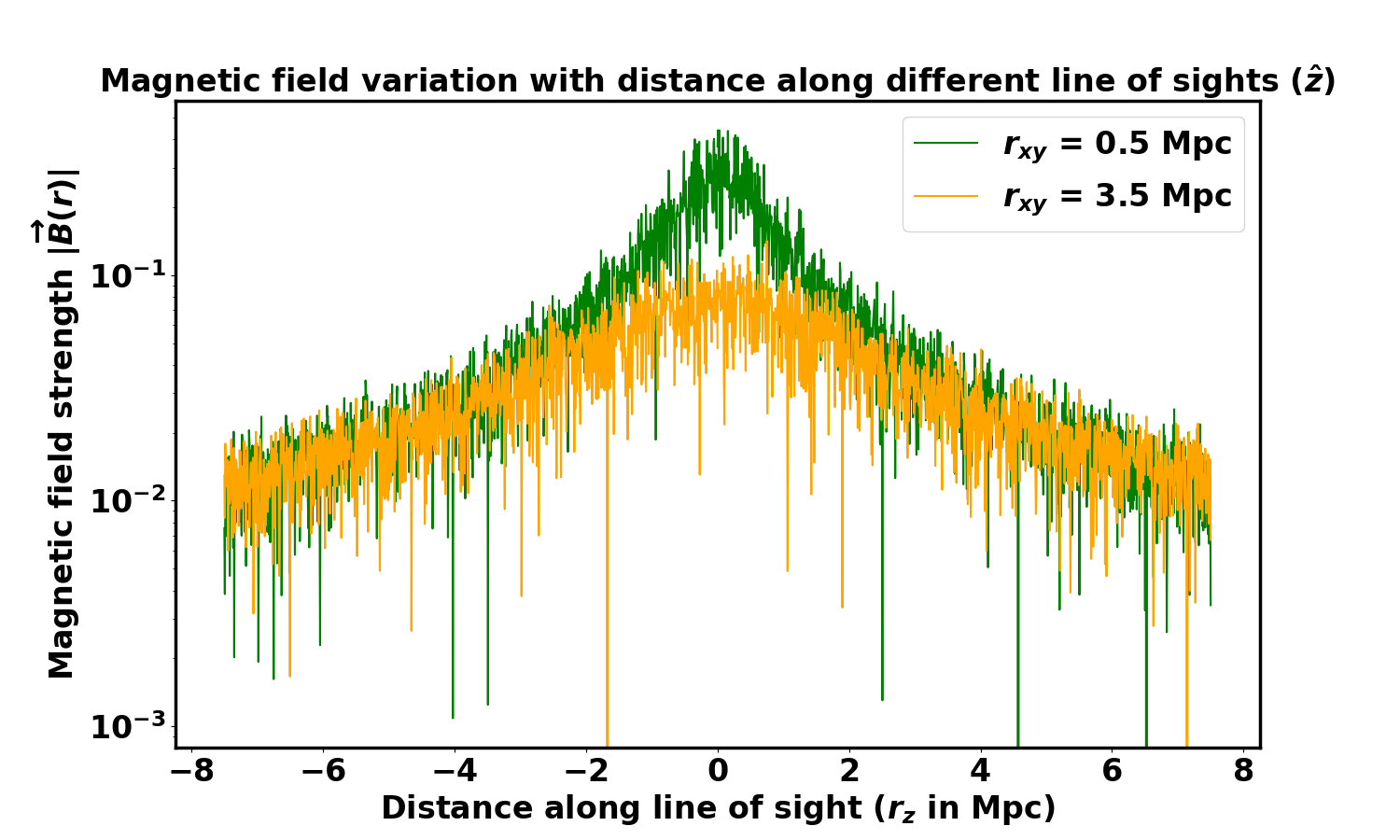}
    \caption{Magnetic field magnitude (with strength inhomogeneity standard deviation being 30\% of the mean radial profile) for different locations along the line of sight for two different projected distances. The length of the setup considered here is 15 Mpc.}
       \label{fig:B_los}
\end{figure}

\begin{figure}[h!]
     \centering
\includegraphics[height=7cm,width=11cm]{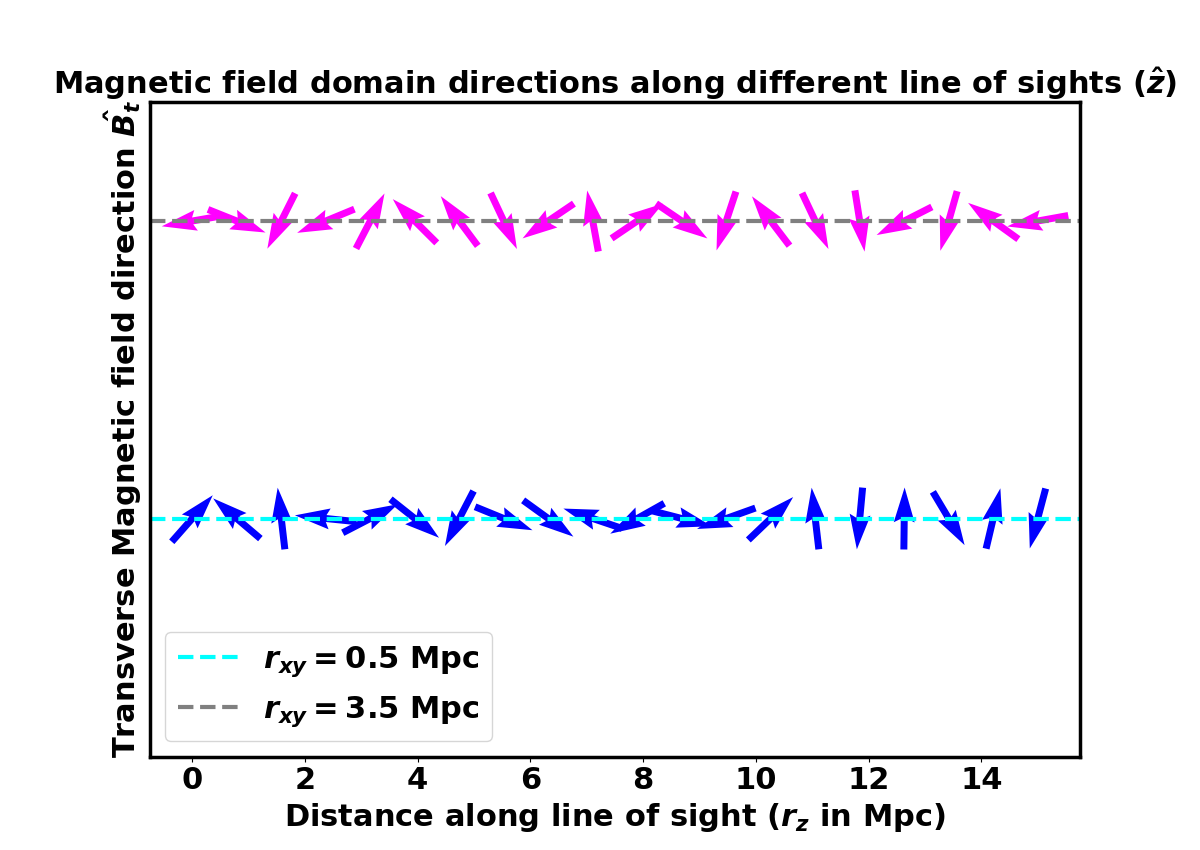}
    \caption{Transverse magnetic field direction for different domains along the line of sight for two different projected distances. The length of the setup considered here is 15 Mpc. We have shown the directions of 20 domains along each line of sight.}
       \label{fig:dir_los}
\end{figure}

 As discussed in the last section, the presence of turbulence in the magnetic field at small scales in galaxy clusters can play a crucial role in estimating the strength of the signal in CMB temperature and polarization anisotropy. 
In our analysis, we model the strength inhomogeneity by a Gaussian distribution over the expected smooth profile and vary the standard deviation of the distribution. The effect of inhomogeneity in the electron density profile for two different lines of sight at different distances from the cluster center is shown in Fig. \ref{fig:ne_los} for a scenario with the standard deviation of the variation $\sigma_e(r)= 0.3\times n_{es}(r)$, where $n_{es}(r)$ denotes the mean electron density as a function of radial separation from the center of the galaxy cluster. Here, $\hat{z}$ represents the line of sight direction, $r_{z}$ represents the distance along the line of sight, and $r_{xy}$ represents the projected distance perpendicular to the line of sight. The horizontal dashed line is the electron density required for resonance condition to be satisfied for $10^{-13}$ eV mass ALPs, which takes place at different longitudinal ($z$) locations for different lines of sight. The magnetic field in the presence of strength inhomogeneity is also shown as a function of distance from the cluster center, with the standard deviation of the variation as $\sigma_B(r)= 0.3\times |\overrightarrow{B(r)}|_s$ (where
$|\overrightarrow{B(r)}|_s$ denotes the smooth radial component of the magnetic field magnitude as a function of radial separation from the center of
the galaxy cluster) in Fig. \ref{fig:B_los}, while the variation of transverse magnetic field direction along different domains (directional inhomogeneity) along the line of sight is shown in Fig. \ref{fig:dir_los}. Here, we assign random directions to the magnetic fields in different domains. In our analysis, we show the impact of the inhomogeneities in electron density and magnetic field for a few different strengths of $\sigma_e$ and $\sigma_B$. 

\section{The Simulation Setup}
\label{sec:method}
The conversion can occur only at resonant locations, which will be determined by the electron density at various locations within the cluster. The ALP distortion signal from this conversion will depend on the length scale of photon-ALP oscillations ($\Delta_{\mathrm{osc}}^{-1}$), the electron density variation scale ($\Delta_{e}^{-1}$), and the coherence scale of the magnetic fields.  Also, turbulence in radial electron density and magnetic field profiles will lead to departure from spherical symmetry of the ALP signal from the cluster. The contribution to the polarization signal will depend on the magnitude of the transverse magnetic field, as well as its direction and coherence scale, while the temperature signal will only depend on the magnitude of the transverse magnetic field at the conversion location.  
We need to take these factors into account when simulating the ALP distortion signal from a cluster. We describe below the simulation setup developed in this work for capturing these effects. 

\subsection{The cluster simulation grid}
\label{ref:cluster_grid}

\begin{figure}
    \centering
    \includegraphics[height=11cm,width=15cm]{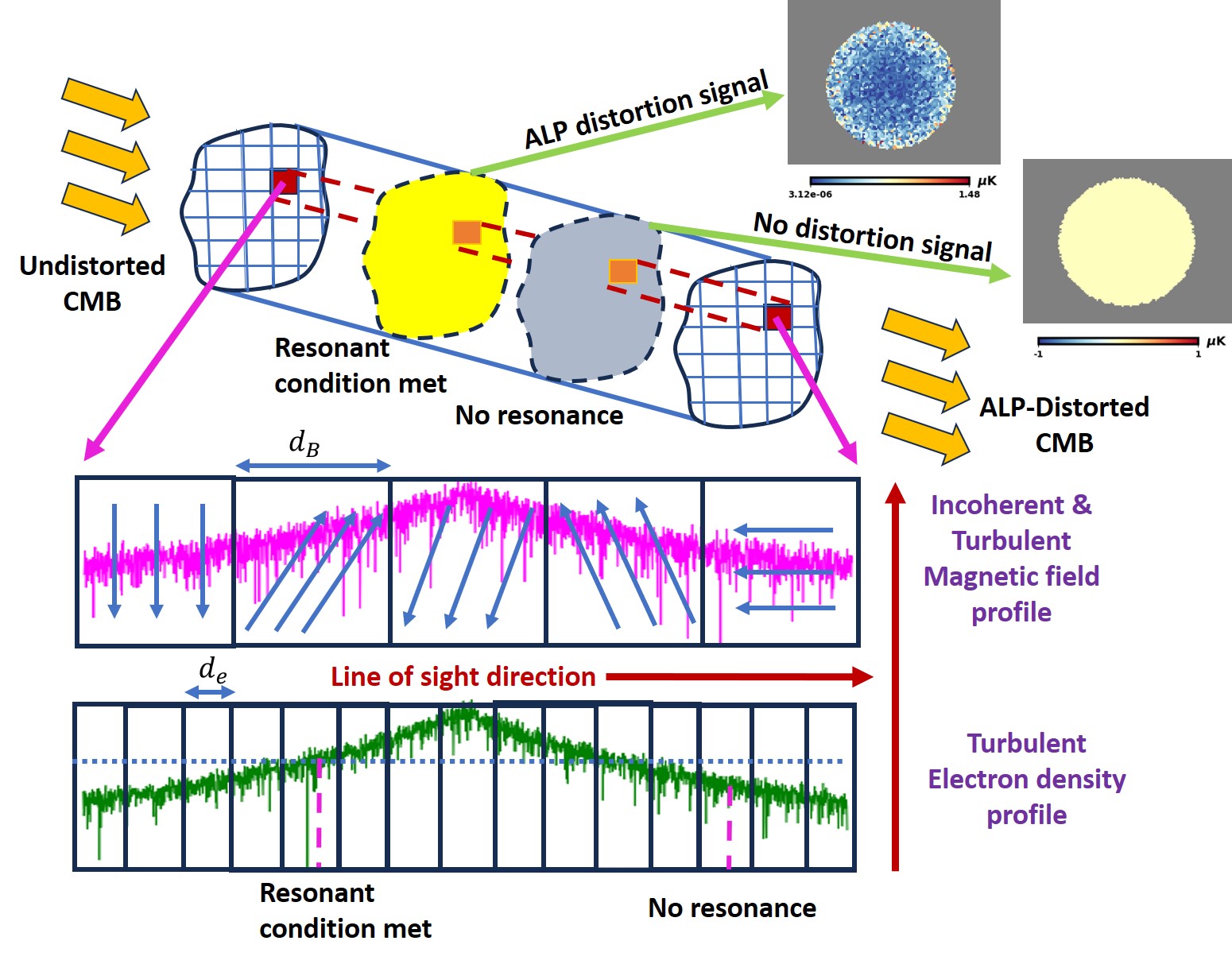}
    \caption{Simulation setup for simulating the ALP signal in a cluster. Electron density and magnetic field strengths can be assigned in grids of different sizes based on the inhomogeneous component. The magnetic field direction can be randomly assigned in those grids or domains.  The simulation setup calculates the ALP signal from the locations where the resonance condition is met based on the electron density of the cluster. The view of a section of the setup where the resonant condition is met shows an ALP signal, while the view of a section where the resonance condition is not met shows  no signal. }
    \label{fig: setup}
\end{figure}
{
We aim to simulate the photon-ALP conversion signal in the presence of electron density and magnetic field in a galaxy cluster along the line of sight. The size of our simulation box should be large enough so that all locations in the cluster where we expect the resonance condition to be satisfied are taken into account.} 
In the simulation setup (see Fig. \ref{fig: setup}), the cluster is projected as a disk in two dimensions. The limits of this disk are set by the angular size of the cluster, which depends on its redshift and its physical extent.  The size of the simulation box should be large enough to be able to probe the ALPs of required masses. Since lower mass ALPs will be formed in outer regions of the cluster, we use a box size of 15 Mpc in our current analysis, which will be large enough to probe the ALPs of low masses. This size refers to the diameter of the projected plane of the cluster and also its length along the line of sight. The pixels occupied by this cluster will correspond to different lines of sight passing through the cluster. We define a cluster center with respect to which the smooth radial profiles are calculated. We define the $x$ and $y$ axes as creating the two-dimensional projected plane containing the cluster pixels, while $z$ axis corresponds to the cluster line of sight. 

 For the case of CMB photons travelling through galaxy clusters, the resonance will be mostly non-adiabatic, for which we expect the grid sizes for electron density variation ($d_e$) to be smaller than oscillation length scale ($l_\mathrm{osc}$). Also, we can set the grid sizes for magnetic field variation in domains of sizes $d_B$.
Thus, each line of sight, which corresponds to the $z$-axis, is subdivided into numerous domains of sizes $d_e$ and $d_B$, where $d_e < l_\mathrm{osc}$. The electron densities will be calculated and assigned in grid sizes of size $d_e$, while magnetic fields will be assigned in grid sizes of size $d_B$. We take $d_B$ to be an integer multiple of $d_e$, so that each grid of size $d_e$ is assigned to some grid of size $d_B$, that occupies the same location. Further, the domains are assigned randomly varying magnetic field direction as depicted by the arrows in Fig. \ref{fig: setup}.

\begin{figure}[h!]
     \centering
\includegraphics[height=6.5cm,width=10cm]{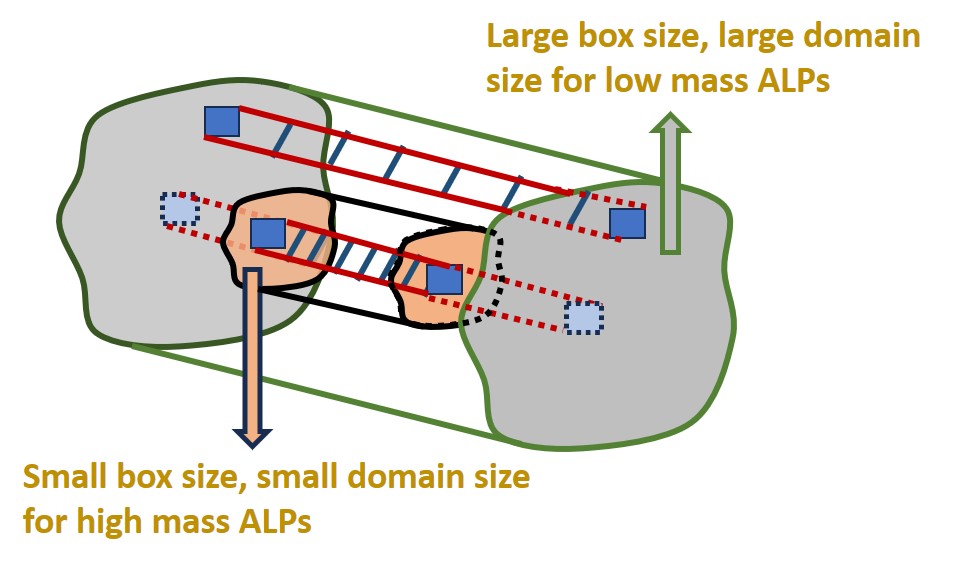}
    \caption{The adjustability of the simulation setup for different ALP masses. The low-mass ALPs are formed in the outer regions of clusters and require a larger box size. The high-mass ALPs are formed in the inner regions of clusters, and the simulation setup can be scaled accordingly with a lower grid size to account for a higher turbulence in the inner regions.}
       \label{fig:scale}
\end{figure}

The grids can be scaled accordingly based on the ALP masses that are being probed (see Fig. \ref{fig:scale}). With high-mass ALPs, we can scale the grid to lower sizes as high-mass ALPs are generally formed in the inner regions of galaxy clusters with higher turbulence. This would allow us to consider the effects of directional and strength inhomogeneities at higher resolutions while also cutting the computational cost. At resonance, the oscillation length scale will mainly depend on the transverse magnetic field $B_t$, and the ALP coupling constant $g_{a\gamma}$ (see Equ. \eqref{eq:lenparams}). For a fixed coupling, the oscillation length scale will decrease with distance from the center as the magnetic field generally increases with distance from the cluster center. The electron density variation scale, thus, needs to be lower for probing high mass ALPs that are formed in the inner regions of clusters. The photon-ALP oscillation scale and the length of the magnetic field coherence scale decides the amount of depolarization of the signal through the conversion probability and the superposition of polarization from multiple resonances. 
Also, the electron density will decide the locations where ALPs are formed. The grid locations and sizes will thus, decide the points where the ALP signal is calculated. 
 
\subsection{Fluctuations in cluster electron density and magnetic field profiles}
\label{sec:turb_grid}
\subsubsection*{Modeling inhomogeneities in electron density} 

{The electron densities in galaxy clusters can be inferred from X-ray and  SZ observations \cite{Birkinshaw_1999,adam2017mapping,govoni2004chandra,vikhlinin2005chandra,Vikhlinin_2006,sunyaev1980microwave,carlstrom2002cosmology,Komatsu_1999,1972CoASP...4..173S,1986rpa..book.....R,sarazin1986x,adam2017mapping,shitanishi2018thermodynamic,komatsu1999submillimeter}.} The observed electron densities are a mix of smooth and fluctuating components varying within the radius of the galaxy cluster. For our analysis, we employ a smooth radially varying modified $\beta$-electron density model, which considers the high electron densities in inner regions, a cusp core with a power law, and the slope at a large radii
\cite{Vikhlinin_2006,mcdonald2013growth,mcdonald2017remarkable,bartalucci2017recovering}:
\begin{equation}
  n_{es}^2 = Z\left[ n_0^2 \frac{(r/r_{c1})^{-\alpha}}{(1 + r^2/r_{c1}^2)^{3\beta - \alpha /2}}\frac{1}{(1 + r^{\gamma}/r_s^{\gamma})^{\epsilon / \gamma}} + \frac{n_{02}^2}{(1 + r^2/r_{c2}^2)^{3\beta_2}}\right].
\label{eq:elec dens}\end{equation}
The electron density values are obtained as the sum of the smooth radial component and a Gaussian turbulent component over it, with the standard deviation denoted by $\sigma_e$, which captures the fluctuation. The electron density at various locations in the grid is given by: 
\begin{equation}
    n_e(x,y,z) = n_{es}(r) +  \mathcal{G}[n_{es}(r), \sigma_e ] \, .
\end{equation}
We assign electron density values to all grid points of size $d_e$ (where $d_e < l_{\mathrm{osc}}$), and this sets the mass of ALPs that can form at a certain location based on the resonant condition. The electron density gradient is determined by finding the derivative between two neighbouring points along the same line of sight, which lies along the $z$-axis. We use the backward difference derivative here, as that is the path being travelled by the photon. This is given as:
\begin{equation}
    \nabla_{z} n_e(x,y,z + d_e) = \frac{n_e(x,y,z + d_e) - n_e(x,y,z)} {d_e} \, .
\end{equation}
This derivative is used to calculate the conversion probability at various resonant locations.

\subsubsection*{Effect of turbulence in magnetic field:}
The magnetic fields in galaxy clusters are possibly a result of the energy injected into the ICM due to the gravitational forces. These fields are amplified by dynamo effects or Active Galactic Nuclei (AGNs) as they inject energy via jets and shocks. This leads to the frozen magnetic fields in the plasma to be compressed and amplified \cite{subramanian2006evolving,donnert2018magnetic,schekochihin2006turbulence,xu2009turbulence,ensslin2006magnetic}.  Based on the observations of low redshift clusters, the magnetic fields in clusters are of the order of 0.1 $\mu \mathrm{G}$, but for cooling-flow clusters, these may reach up to 30-50 $\mu \mathrm{G}$ \cite{GOVONI_2004,carilli2002cluster,Ferrari_2008}. The magnetic field profiles are quite complex in clusters, with a mix of coherent and turbulent components. For our analysis, we employ a smoothly radially varying magnetic field profile that is motivated from observations of low redshift clusters \cite{Carilli_2004,bonafede2010galaxy,bohringer2016cosmic}:
 \begin{equation}
   B_s(r) = B_0 r^{-s},
\label{eq:mag prof}
\end{equation}
where $r$ is the distance of a grid point with domain size $d_B$ from the cluster center. Strength inhomogeneity is injected at various locations in the grid as a Gaussian variation with a standard deviation denoted by $\sigma_B$ over the smooth magnetic field amplitude as 
\begin{equation}
    B(x,y,z) = B_{s}(r) +  \mathcal{G}[B_{s}(r), \sigma_B ] \, .
\end{equation}
{The components of magnetic field amplitude can be decomposed into longitudinal and transverse components with respect to the line of sight of an observer.} 

The longitudinal magnetic field along the line of sight can be studied using Faraday Rotation measurements, a phenomenon that leads to a rotation of the photon polarization plane as it travels through the ionized plasma. This is a result of the preference of electrons in a medium to rotate in a certain direction depending on the magnetic field direction \cite{GOVONI_2004,murgia2004magnetic,bohringer2016cosmic,Ferrari_2008,clarke2001new,eilek2002magnetic,bonafede2010galaxy}. The transverse component of the magnetic field along the line of sight ($B_t = \sqrt{B_x^2 + B_y^2}$) can be obtained using synchrotron observations, which takes place when relativistic electrons are accelerated in a transverse magnetic field \cite{GOVONI_2004,bohringer2016cosmic,clarke2001new,Ferrari_2008,1986rpa..book.....R,condon2016essential}. 
This is the component responsible for the ALP conversion along the cluster line of sight. The longitudinal component $B_z$ will thus not be impacting the ALP distortion signal. The signals from both the Faraday measurements, as well as the synchrotron emission, depend on the coherence scale of the magnetic fields. This dependence on the coherence scale of these signals enables us to separate the magnetic field into coherent and turbulent components \cite{schekochihin2006turbulence,lee2019anisotropic,basu2019depth,pfrommer2010detecting}. On large scales, it is the coherent component of the magnetic field due to the bulk motion of charges that determines the signals, while the turbulent component affects the signals on small scales due to the random motion of charges locally. This motivates the use of an observationally motivated coherent magnetic field with Gaussian fluctuations about the mean. 

The conversion of photons to ALPs will reduce the CMB brightness along the cluster line of sight, which lies along the $z-$axis in our setup. The strength of this effect will depend on the transverse magnetic field magnitude. 
It is the variation of the $B_x$ and $B_y$ components that will lead to the depolarization of the ALP distortion signal as the CMB photon travels along the line of sight through the cluster, and there are multiple resonant locations. We set the coherence scale of the magnetic field by changing the domain size $d_B$. The change in magnetic field direction is brought by randomly varying the magnetic field components along the three axes ($B_x, \, B_y$ and $B_z$) in those domains. This requires us to define the unit vectors along the three axes as follows 
\begin{equation}
     \hat{B} \cdot \hat{z} = \cos{\theta_z} , \, \, 
     \hat{B} \cdot \hat{x} = \sin{\theta_z} \cos{\phi_x}, \, \, 
     \hat{B} \cdot \hat{y} = \sin{\theta_z} \sin{\phi_x}, \, \, 
\end{equation}
where $\hat{B}$ refers to the unit vector along the magnetic field direction in the grid, $\theta_z$ is the angle made by the magnetic field vector with the $z$ axis, while $\phi_x$ is the angle made by the projection of the magnetic field vector on the $x-y$ plane with the $x-$axis.

The electron density profile will determine the mass of ALP that can be produced at a given location, while the transverse magnetic field magnitude and the electron density gradient determine the strength of this conversion. The profiles can be broken down to radially smooth and turbulent components, with the distance from cluster center determining the smooth profiles. The turbulent component will determine the deviation from a smooth profile at different domain points. The amount of strength inhomogeneity is taken as the Gaussian deviation over the smooth profile with the standard deviation defined with respect to the smooth profile magnitude at the corresponding domain point. The realization grids allow us to vary the strength inhomogeneity in electron density and magnetic field magnitudes, as well as change the direction of the magnetic field at different points. 

The Gaussian turbulence approximation fails on small non-linear scales or when the AGN or cluster merger induced shocks produce anisotropic non-Gaussian statistics.  However, it is a valid model for relaxed clusters with isotropic turbulence. It is also a good assumption on large scales when the small scale non-linear effects are averaged out  \cite{subramanian2006evolving,xu2009turbulence,schekochihin2006turbulence}. In this analysis, we focus on Gaussian fluctuations in the electron density and magnetic field. But our simulation setup can incorporate any other distribution as well.

\section{ALP distortion in intensity and polarization}
\label{sec:powspectra}
The ALP signal can be probed either using temperature or polarization of the CMB along the cluster line of sight \cite{Austermann_2012,2020_planck}. The temperature map will contain information of the electron density and magnetic field magnitudes, while the magnetic field direction at the conversion locations is crucial in determining the polarization signal. The ALP distortion signal in unpolarized intensity is given as a sum of the loss in CMB intensity due to photon-ALP conversion. {This simplifies to the sum of conversion probabilities at the leading order} at different locations along the line of sight  
\begin{equation}
    \Delta I_T^{\alpha}(\nu) = \frac{\pi}{2} \sum_{i = 1}^{N}  \gamma_{\mathrm{ad},i} \, I_0 (\nu) \, ,
   \label{eq: tempalp}
\end{equation}
where $I_0 (\nu)$ refers to the CMB black-body intensity and $N$ is the number of resonances along the line of sight. The ALP temperature signal generally increases in the outer regions of the signal disk and then decays down to zero as the ALP signal is limited to the extent of the disk \cite{Mukherjee_2020,Mehta:2024sye,Mehta:2024wfo}. 

The superposition of polarized photons from different locations along the line of sight will determine the observed polarization corresponding to the ALP distortion signal. The polarization signal is decomposed to the Q and U Stokes parameters \cite{ghatak2009optics,Austermann_2012,2020_planck}. This decomposition depends on the magnetic field direction at various resonances along the line of sight. If the transverse magnetic field makes an angle $\phi_x$ with the $x$-axis of our grid, the polarized intensity can be split to Q and U maps as: 
\begin{equation}
    \Delta I_Q^{\alpha}(\nu) =  \frac{\pi}{2} \sum_{i = 1}^{N}  \gamma_{\mathrm{ad},i} (\cos{2\phi_x})_i \, I_0 (\nu) \, ,
   \label{eq: tempalp}
\end{equation}

\begin{equation}
    \Delta I_U^{\alpha}(\nu) =  \frac{\pi}{2} \sum_{i = 1}^{N}  \gamma_{\mathrm{ad},i} (\sin{2\phi_x})_i \, I_0 (\nu) \, ,
   \label{eq: tempalp}
\end{equation}
 The polarized intensity is then given as:
\begin{equation}
    \Delta I_{P}^{\alpha} (\nu) =  \sqrt{\Delta I_Q^{\alpha}(\nu)^2 + \Delta I_U^{\alpha}(\nu)^2} \, .
   \label{eq: tempalp}
\end{equation}
In order to convert the intensity values to the CMB temperature units, we use: 
\begin{equation}
    \Delta T^{\alpha} = \left(\frac{dI}{dT}\right)_{\mathrm{cmb}}^{-1} \Delta I^{\alpha} \, ,
\end{equation}
where the derivative is taken for the CMB blackbody spectrum.
If the magnetic field lines are all ordered in a particular direction at all locations where resonant conversion occurs, the polarized signal will be boosted and match the temperature signal. For the case of random magnetic field direction at different conversion locations, the polarization signal will be suppressed due to depolarization with a decrease in the coherence length of the ordered magnetic field  \cite{subramanian2006evolving}.

The presence of turbulence will affect the temperature and polarization signals from the cluster. The presence of magnetic field turbulence will change the amplitude of the signal from different lines of sight depending on the turbulent magnetic field profile at the conversion locations. The fluctuations in the signal will be more apparent due to the change in the magnetic field. However spatial extent or the spatial features of the ALP signal disk would not be affected and will be similar to the case for a smooth radially varying magnetic field profile. 

The presence of turbulence in electron density will change not only the amplitude of the signal but also the size and shape of the signal disk due to a change in resonant locations within the cluster. {The circular shape as well as the symmetry of the signal disk will be broken down in this case,} accompanied with an increase in the thickness of the conversion shell, which would increase the size of the signal disk. Also, the magnitude of the signal along different lines of sight will change as the effective conversion probability changes due to a change in resonant locations as well as a change in the electron density gradient.   
The difference in power associated with the temperature and polarization spectra depends highly on the magnetic field coherence scale, with a large suppression of power in polarization for a lower coherence scale, as more number of domains will not be in phase directionally. The power in temperature depends only on the magnitude of the transverse magnetic field. Thus, although these signals will be different, their origin remains the same, and hence, the two signals will be correlated based on the turbulence in profiles. This can be studied based on the statistics of the cluster region in both temperature and polarization. Also, the signals will be spectrally correlated, independent of the spatial fluctuations. These aspects are considered in the next section.

\section{Results}
\label{sec:results}
In this section, we calculate the ALP distortion power spectra in CMB temperature and polarization field for varying scenarios of electron density and magnetic field models. The simulated cluster is considered at a redshift of $z=0.15$ with a pixel resolution for the HEALPix NSIDE = 2048 \cite{2005ApJ...622..759G,Zonca2019}. The computation times for individual simulation case runs are around 21 core-hours on a machine with a clock speed of 3.6 GHz for a simulation of a galaxy cluster consisting of approximately 700 HEALPix pixels with 15000 domains along a line of sight. The simulation setup can work for any ALP masses, coupling strength, and cluster source redshift. 
  
 We have set the variation scale of electron density and coherence scale of magnetic field to $d_e = 100$ pc and $d_B = 100$ pc respectively in our analysis, except for the cases of variation of coherence scale, and for high-mass ALPs, as the photon-ALP conversion takes place near the core of the cluster where the electron density is high, {we use a grid sizes of $d_e = 1$ pc and $d_B = 1$ pc, to be able to resolve the inner part of the cluster in order to capture the inhomogeneities in the electron density and magnetic field at small scales.} 

We list below different physical scenarios of variation in electron density and magnetic field considered in this paper to show its impact on the ALP signal:
 \begin{itemize}
     \item 
     \textbf{Change in magnetic field domain size:} Considers the effect of depolarization of the ALP signal due to varying domain sizes of galaxy cluster magnetic field with a radial homogeneous profile of the magnetic field strength.
     \item \textbf{Inhomogeneity in magnetic field strength:} Considers the effect of fluctuation in the magnetic field strength while keeping the same magnetic field domain size.
     \item \textbf{Turbulence in magnetic field:} Considers the effect of varying magnetic field strength as well as the domain size on the strength of the ALP signal.
     \item \textbf{Turbulence in electron density:} Considers the effect of varying electron density profile and strength on the ALP signal.
     \item \textbf{Inhomogeneity in magnetic field and electron density strength:} Considers the effect of fluctuation in both magnetic field and electron density strength on the spatial features of the ALP signal. 
 \end{itemize}
 We break down our analysis into all these different cases to isolate the impact of these different physical effects on the ALP signal. Our fiducial model considers ALPs of masses $m_a= 10^{-13}$ eV and coupling strength $g_{a \gamma}= 10^{-12}$ GeV$^{-1}$. We also show a result from our simulation technique for a higher ALP mass in the appendix \ref{sec:high mass}. A high ALP mass simulation requires high resolution to probe the inner regions of the cluster to take into account the contributions from turbulence in those regions. We have shown that our simulation technique can capture this by adaptively changing the resolution.
 
 \subsection{Observables on the Microwave Sky}
 \subsubsection{Spectral correlation of the TP cross spectrum}
\label{sec:TPspec}

\begin{figure}[h!]
     \centering
\includegraphics[height=7cm,width=11cm]{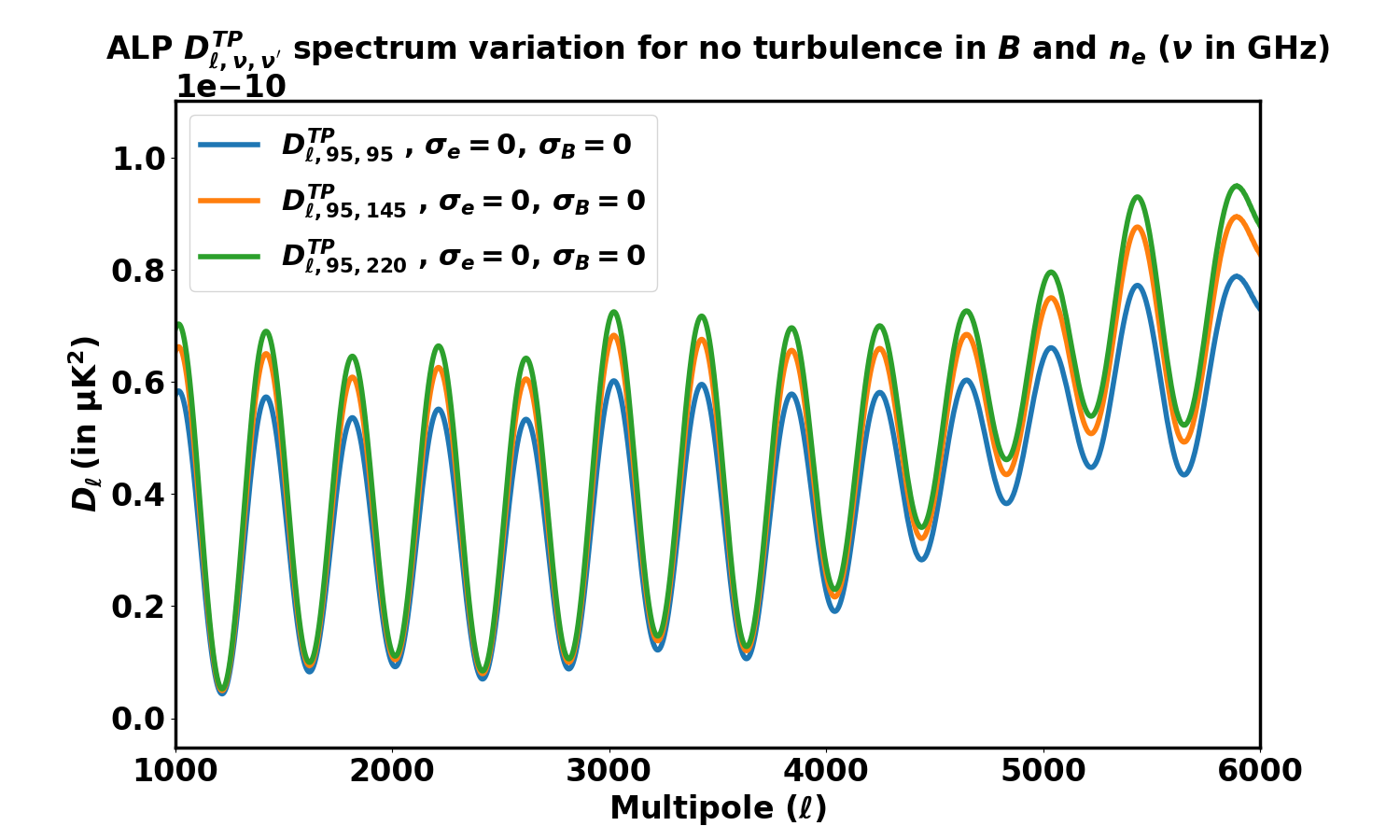}
    \caption{Variation in ALP TP cross spectrum for different sets of frequencies for the case of no strength inhomogeneity.}
       \label{fig:TPf_0}
\end{figure}

\begin{figure}[h!]
     \centering
\includegraphics[height=7cm,width=11cm]{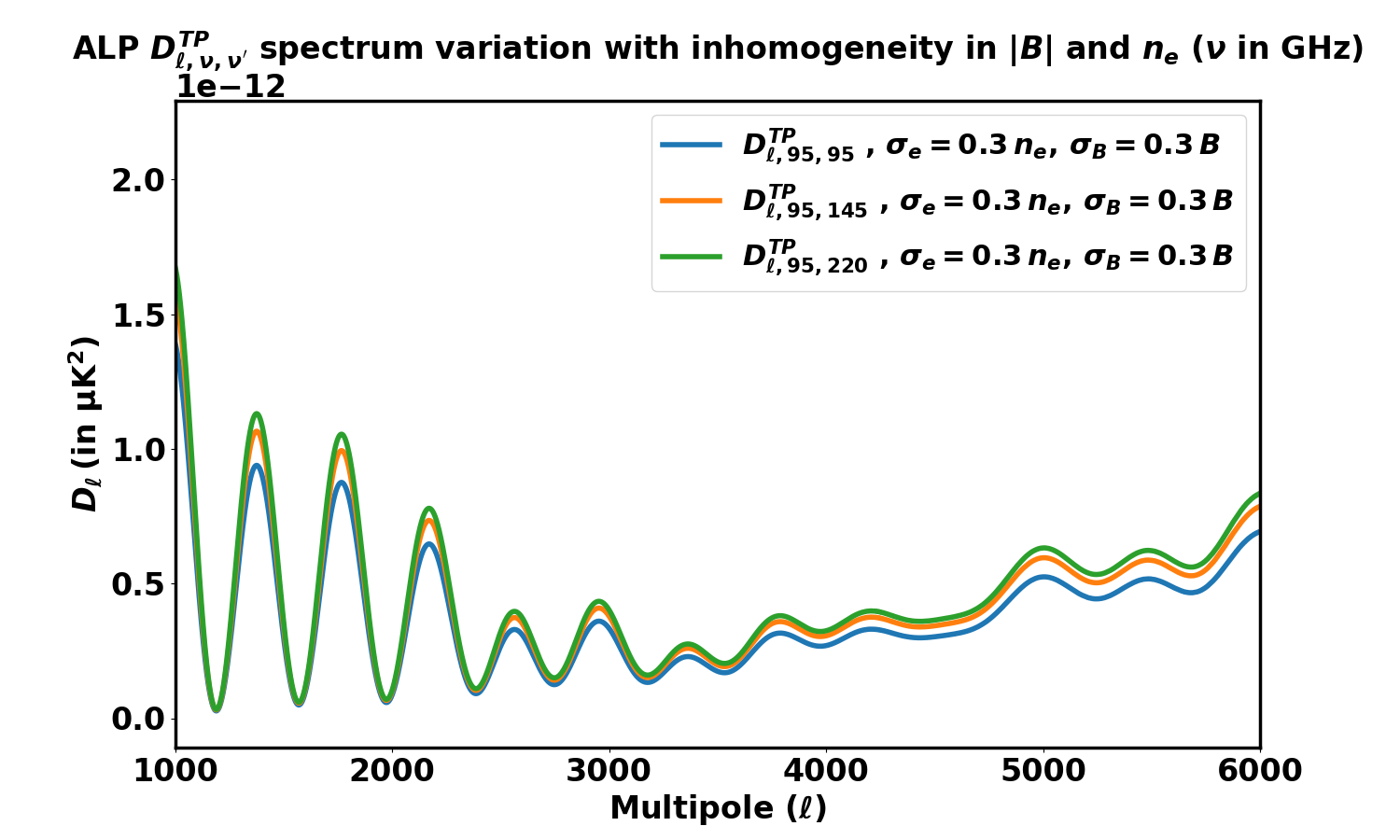}
    \caption{Variation in ALP TP cross spectrum for different sets of frequencies for the case of high strength inhomogeneity in electron density and magnetic field (standard deviation being 30\% of the mean).}
       \label{fig:TPf_3}
\end{figure}
{The photon-ALP conversion results in the decrease of CMB temperature due to a loss in its intensity, as well as a polarization of the photons. Thus, the ALP temperature and polarization signals are expected to show a correlation, depending on the magnetic field direction in different domains, where the resonant conversions can take place. Thus, a temperature-polarization (TP) cross spectrum will contain information about the correlation between temperature and polarization signals at different angular scales, as well as the electron density and magnetic field profiles of the cluster, which determine the signal. A negative value of the cross spectrum at some multipoles will suggest an anti-correlation between the temperature and polarization signals at those angular scales.

Not only can the TP cross spectrum be used to study the correlation between temperature and polarization signals, but it will also follow a well-defined spectral behaviour if the temperature and polarization signals are considered at different frequencies, in contrast to the CMB TP cross spectrum, which is independent of frequency.}
The ALP TP cross spectrum will follow a distinct scaling with frequency based on the ALP distortion spectral variation if the temperature and polarization signals are considered at different frequencies. The ALP distortion signal in both temperature and polarization varies as:
\begin{equation}
    \Delta I^{\alpha} (\nu) \propto \nu I_0(\nu) \, ,
\end{equation}
where $\Delta I^{\alpha}$ is the intensity of the ALP distortion signal, and $I_{0}(\nu)$ is the CMB black-body intensity at the frequency $\nu$. Thus, the TP cross-spectrum for a set of different frequencies will be correlated at all multipoles. This is shown for the case of no strength inhomogeneity in Fig. \ref{fig:TPf_0}, and for the case of high strength inhomogeneity (standard deviation being 30\% of the mean) in both the magnetic field and electron density profiles in Fig. \ref{fig:TPf_3}. The spectra for the two cases vary by about two orders of magnitude, but the spectral scaling in each case is independent of the spatial profile of the signal and, hence, the associated turbulence.   This scaling serves as an independent way to search for ALPs, as it considers the impact of the ALP distortion signal in both the CMB temperature and polarization while also being independent of the uncertainties in cluster profiles. 
The shape of the power spectra is determined by the spatial profile of the ALP signal, which, as can be seen, varies for the homogeneous and inhomogeneous cases at different multipoles. The oscillations are also a result of the shape and size of the cluster region on the sky, the geometry of which determines the masking kernel for a window function with pixels in the cluster region as one and the rest being zero \cite{Hivon_2002}. Thus, these oscillations show up in the power spectra of the upcoming sections as well.

\subsubsection{Induced non-Gaussianity in the photon-ALPs conversion}
\label{sec:gauss}

\begin{figure}[h!]
     \centering
\includegraphics[height=7cm,width=11cm]{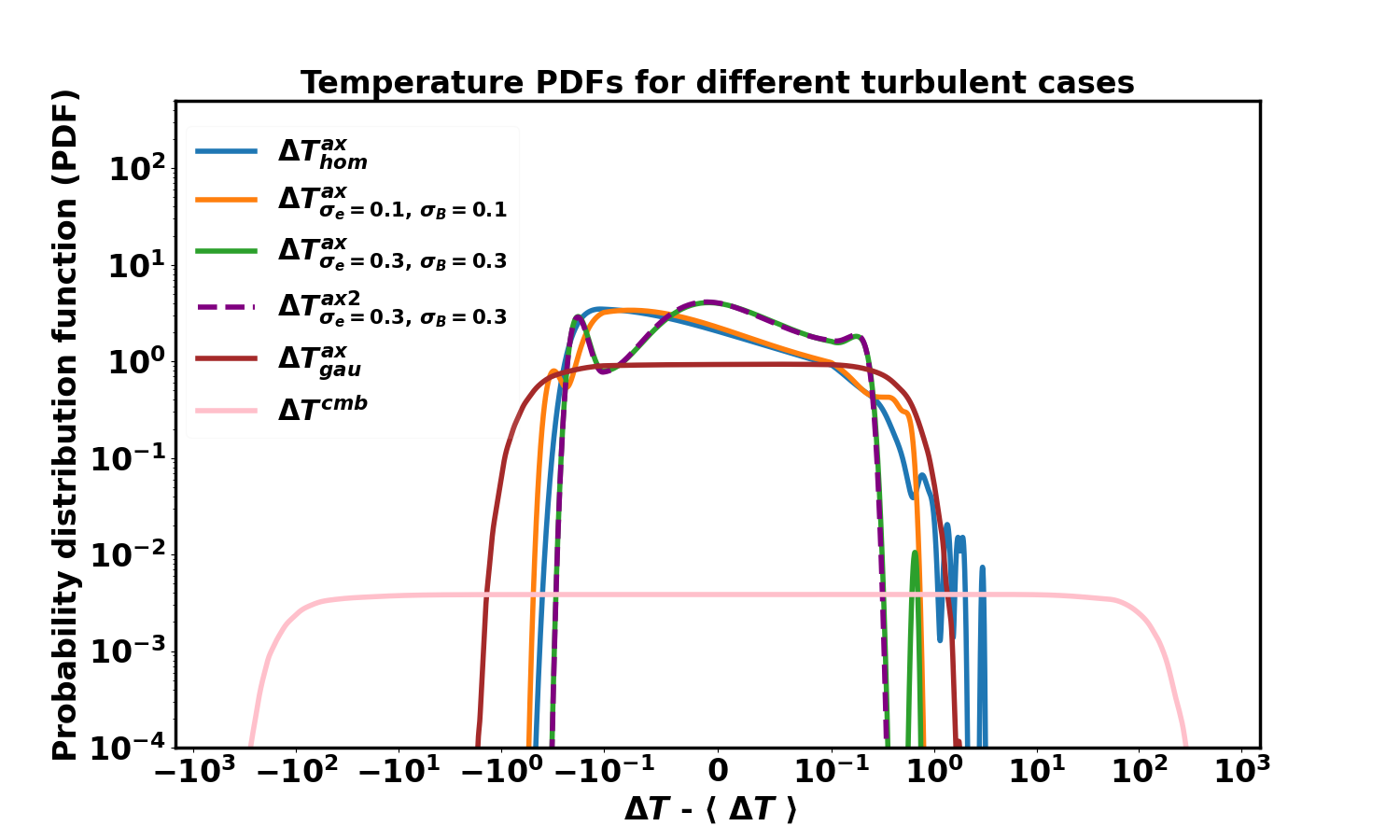}
    \caption{Probability distribution functions of the temperature signal amplitudes with their means subtracted. The percentage refers to the percentage of strength inhomogeneity in electron density and temperature. The curve with the label "ax2" refers to a different realization with the corresponding percentage of strength inhomogeneity in electron density and magnetic field. Here $g_{a\gamma} = 10^{-11} \, \mathrm{GeV^{-1}}$.}
       \label{fig:gaussT}
\end{figure}
\begin{figure}[h!]
     \centering
\includegraphics[height=7cm,width=11cm]{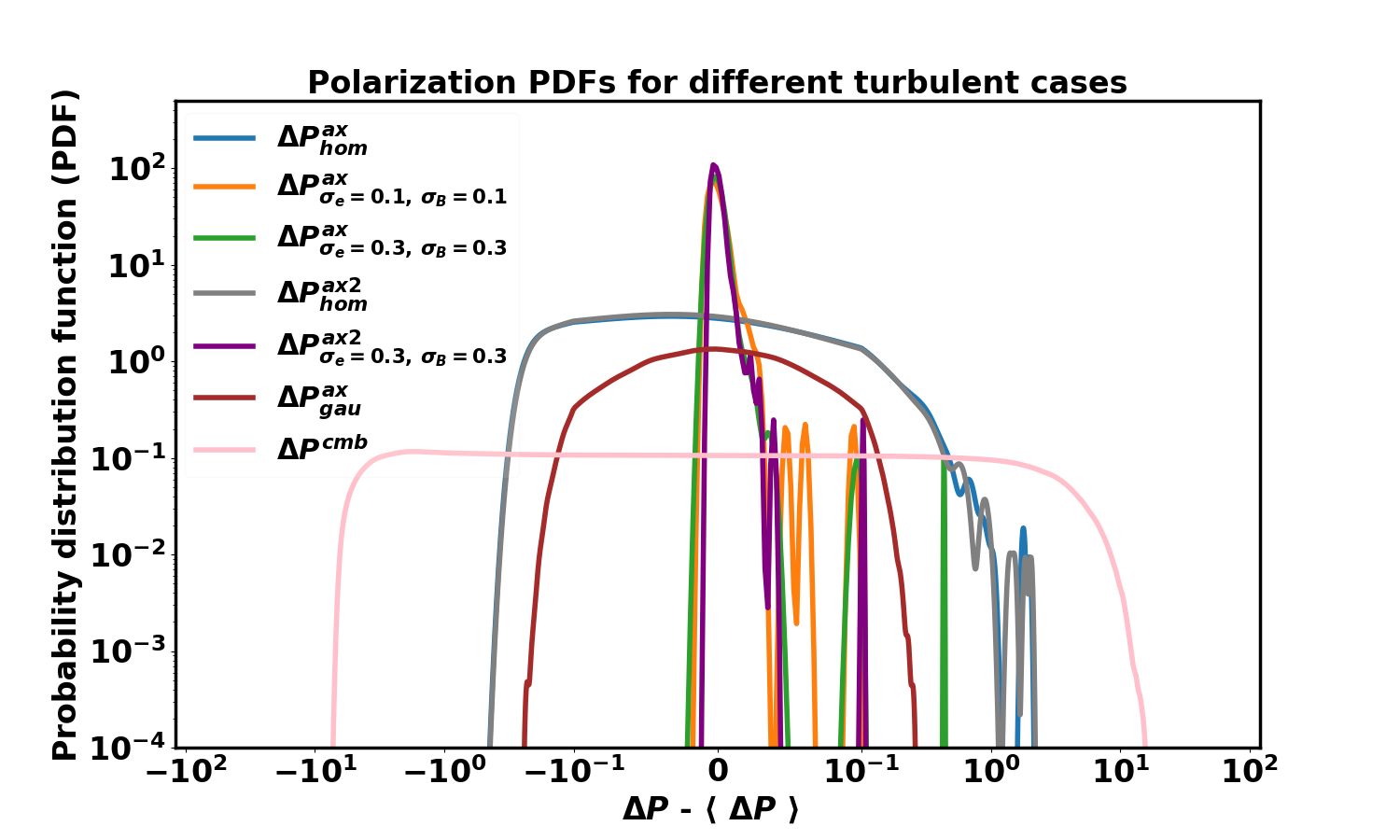}
    \caption{Probability distribution functions of  the polarization signal amplitudes with their means subtracted. The percentage refers to the percentage of strength inhomogeneity in electron density and temperature. The curves with labels "ax2" refer to a different realization with the corresponding percentage of strength inhomogeneity in electron density and magnetic field. Here $g_{a\gamma} = 10^{-11} \, \mathrm{GeV^{-1}}$. }
       \label{fig:gaussP}
\end{figure}

The ALP distortion signal in temperature is characterized by increased strength in the outer regions, followed by a sharp drop. This profile is determined by the electron densities and magnetic fields within the cluster. The polarization signal also depends on the magnetic field directions in different domains, which, for a random distribution, leads to depolarization. The statistics of the ALP signal will thus portray deviations from Gaussianity due to an asymmetry of the signal distribution about the mean value. 

The statistics of the ALP signal can help in separating the ALP signal from CMB. The CMB primary anisotropy is well characterized by Gaussian fluctuations in temperature and polarization \cite{ade2016planck,hanson2009estimators,challinor2012cmb,liguori2006testing,aghanim1999searching}. In Fig. \ref{fig:gaussT}, we show the probability distribution functions (PDFs) of the ALP signal for the case of various levels of inhomogeneity (standard deviations being 0\% (hom), 10\% and 30\% of the mean respectively) in both magnetic field and electron density, with the mean of the histograms subtracted in each case. The CMB follows an almost Gaussian distribution, while the Gaussian ALP  shown refers to the PDF of 100 Gaussian ALP temperature realizations with the same power spectrum as in the absence of inhomogeneity. The labels with "ax2" refer to a different seed of the ALP signal. The ALP temperature signal for the case of no strength inhomogeneity shows a highly non-Gaussian distribution, while it shows a reduction in high signals for the case of high inhomogeneity as fluctuations lead to a higher number of resonances, but these are much weaker as compared to the homogeneous case.

{The polarization signal will also portray non-Gaussianity (see Fig. \ref{fig:gaussP}), which varies with the amount of inhomogeneity. The CMB follows a Gaussian distribution, while the Gaussian ALP shown refers to the PDF of 100 Gaussian ALP polarization realizations with the same power spectrum as for the homogeneous case. The non-Gaussianity of the polarization signal can increase or decrease with inhomogeneity, depending on the number of resonant conversions, the strength of those conversions, and the amount of depolarization along individual lines of sight. For the particular cluster shown, the non-Gaussianity increases with inhomogeneity in the profiles as the distribution becomes highly peaked with a heavy tail distribution.

The non-Gaussian aspect of the ALP signal opens up a new window to probe ALPs in the face of suppressed polarization signals. 
The non-Gaussianity of the region will vary from cluster to cluster depending on the turbulence in the region if ALPs exist, For a particular mass ALPs, the distribution of the observed fluctuations will be determined by the ALP coupling and turbulence in the region. In the absence of such a non-Gaussian signal, the statistics will be more or less determined by the variance of the Gaussian CMB fluctuations. This variation in statistics can be used to break the degeneracy between the suppression of signals due to turbulence, as compared to the case of a weak photon-ALP coupling.}

\subsection{Incoherence in magnetic field}
\label{sec:order}
\begin{figure}[h!]
     \centering
\includegraphics[height=13cm,width=13cm]{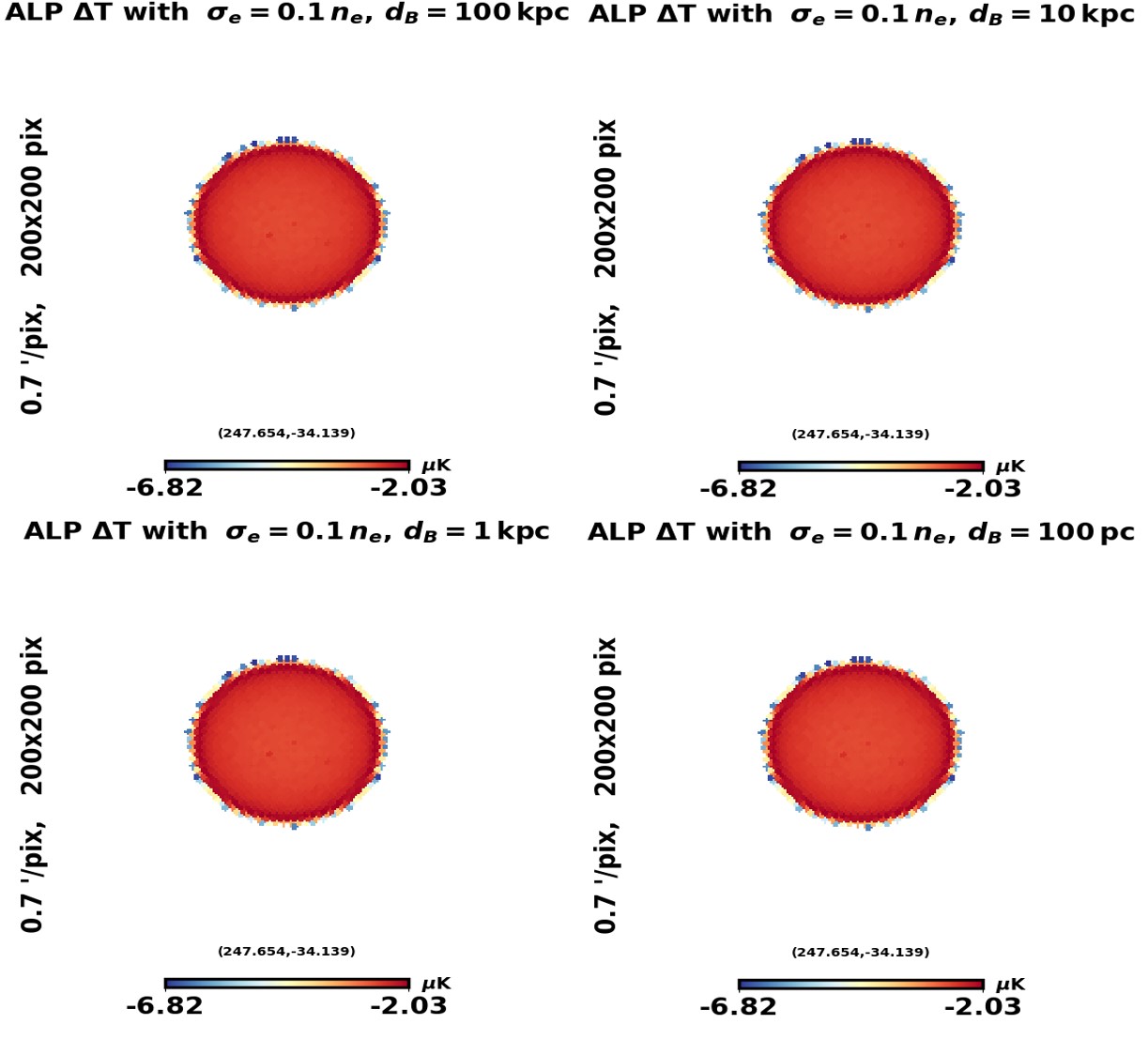}
    \caption{There is no variation in ALP temperature signal with change in magnetic field domain size $d_B$. The values are in log scale.}
       \label{fig:coherT}
\end{figure}

\begin{figure}[h!]
     \centering
\includegraphics[height=13cm,width=13cm]{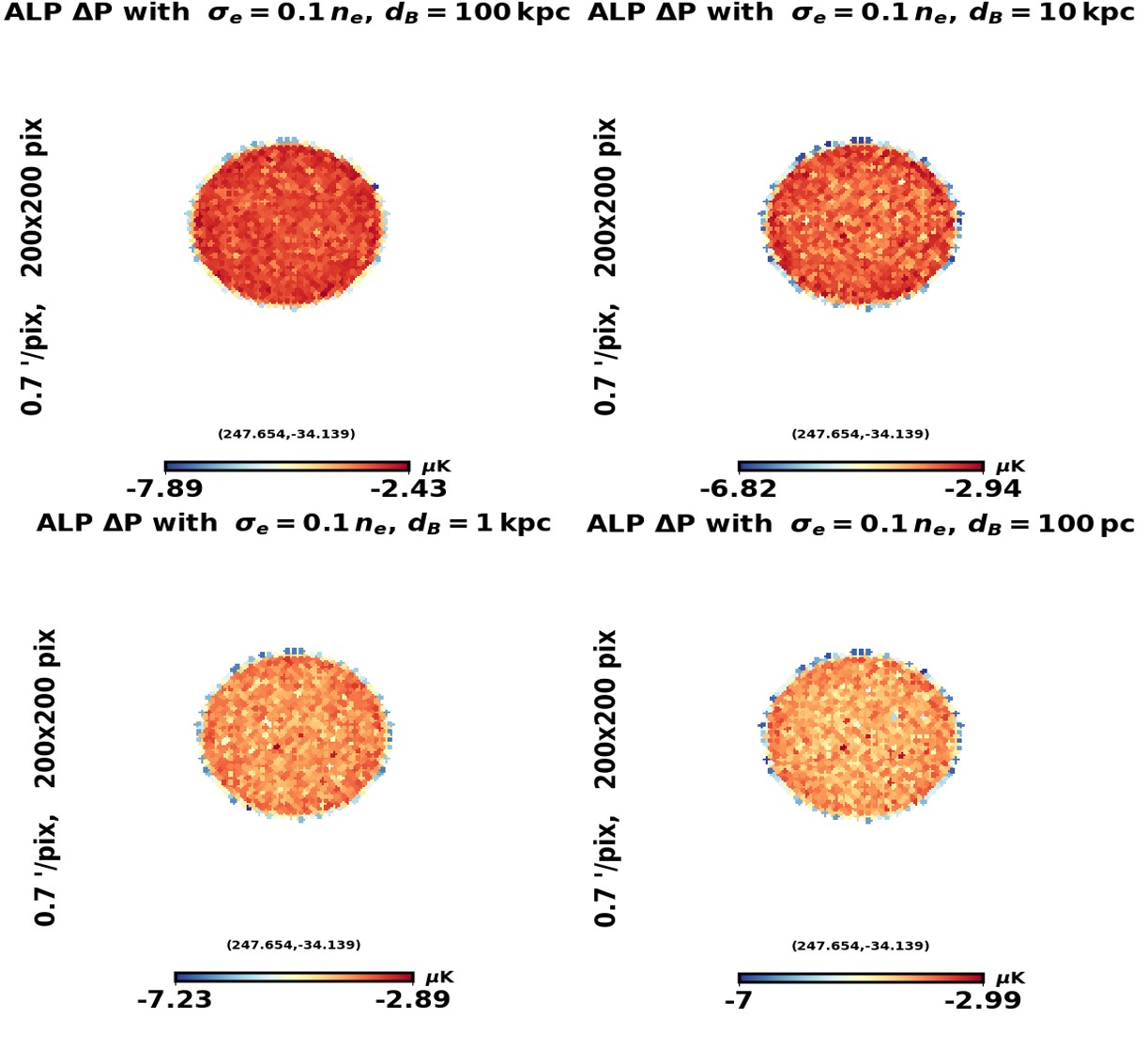}
    \caption{Variation in ALP polarization signal with change in magnetic field domain size $d_B$. The spatial features of the polarization signal change depending on the relative directions of different domains. The magnitude of the polarization signal decreases as the number of domains belonging to the conversion location increases. The values are in log scale.}
       \label{fig:coher}
\end{figure}

\begin{figure}[h!]
     \centering
\includegraphics[height=7cm,width=11cm]{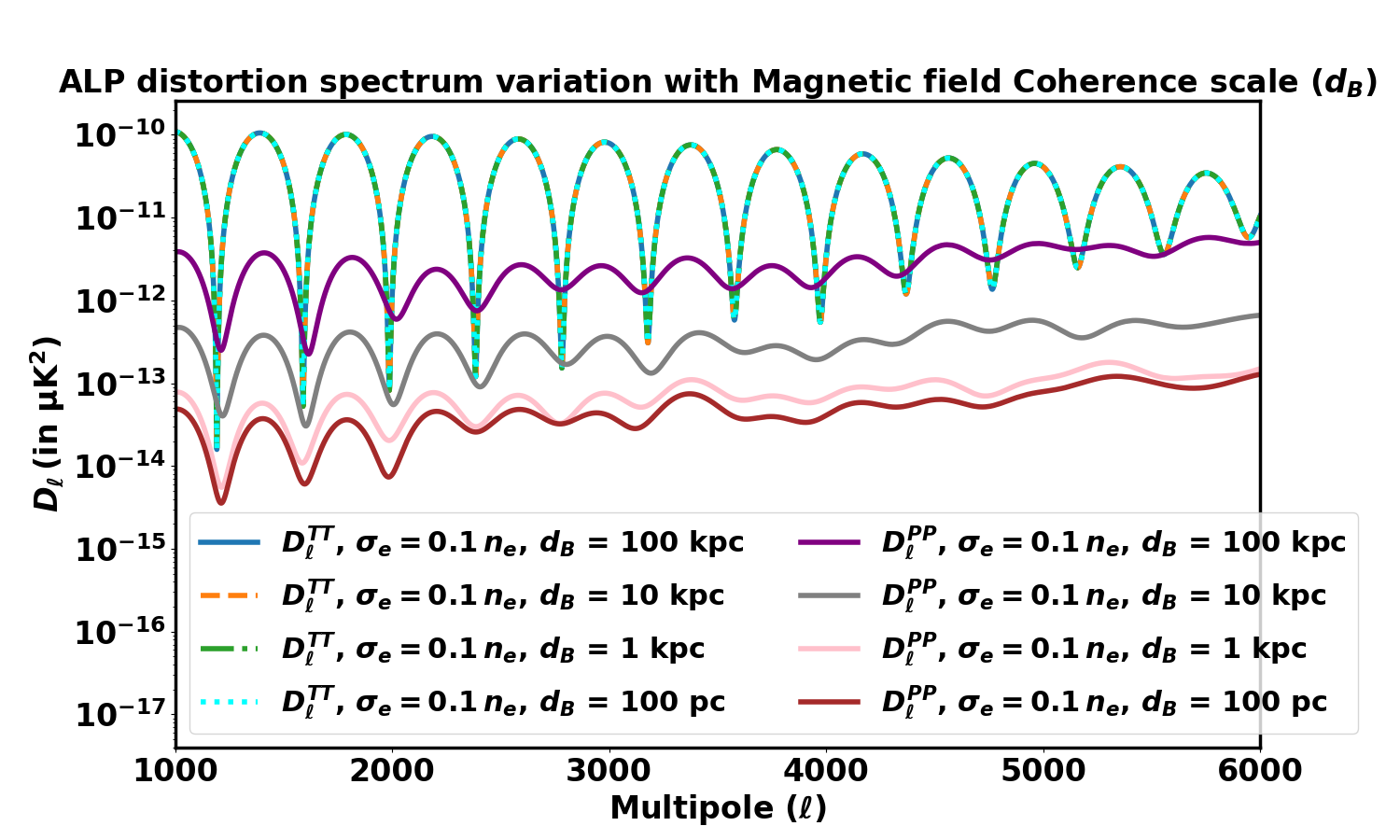}
    \caption{Variation in ALP temperature (TT) and polarization (PP) spectra with change in magnetic field domain size. The amplitude of the PP spectrum decreases as the number of domains belonging to the conversion location increases, which happens for lower domain sizes. The TT spectra for different cases overlap and are indistinguishable.}
       \label{fig:coher_TT}
\end{figure}
\begin{figure}[h!]
     \centering
\includegraphics[height=7cm,width=11cm]{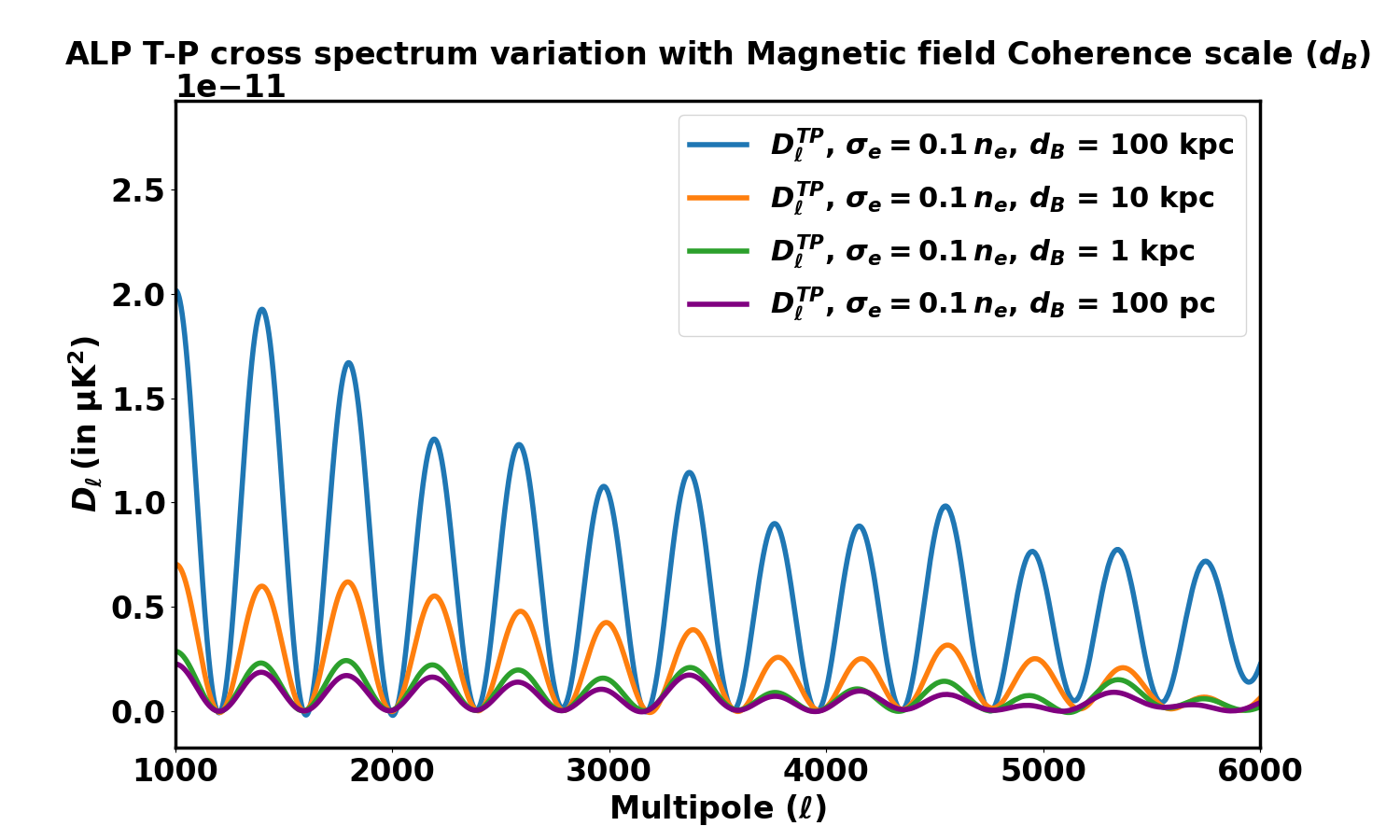}
    \caption{Variation in ALP TP cross spectrum with change in magnetic field domain size. The correlation is higher for higher domain size. }
       \label{fig:coher_TP}
\end{figure}

\begin{figure}[h!]
     \centering
\includegraphics[height=7cm,width=11cm]{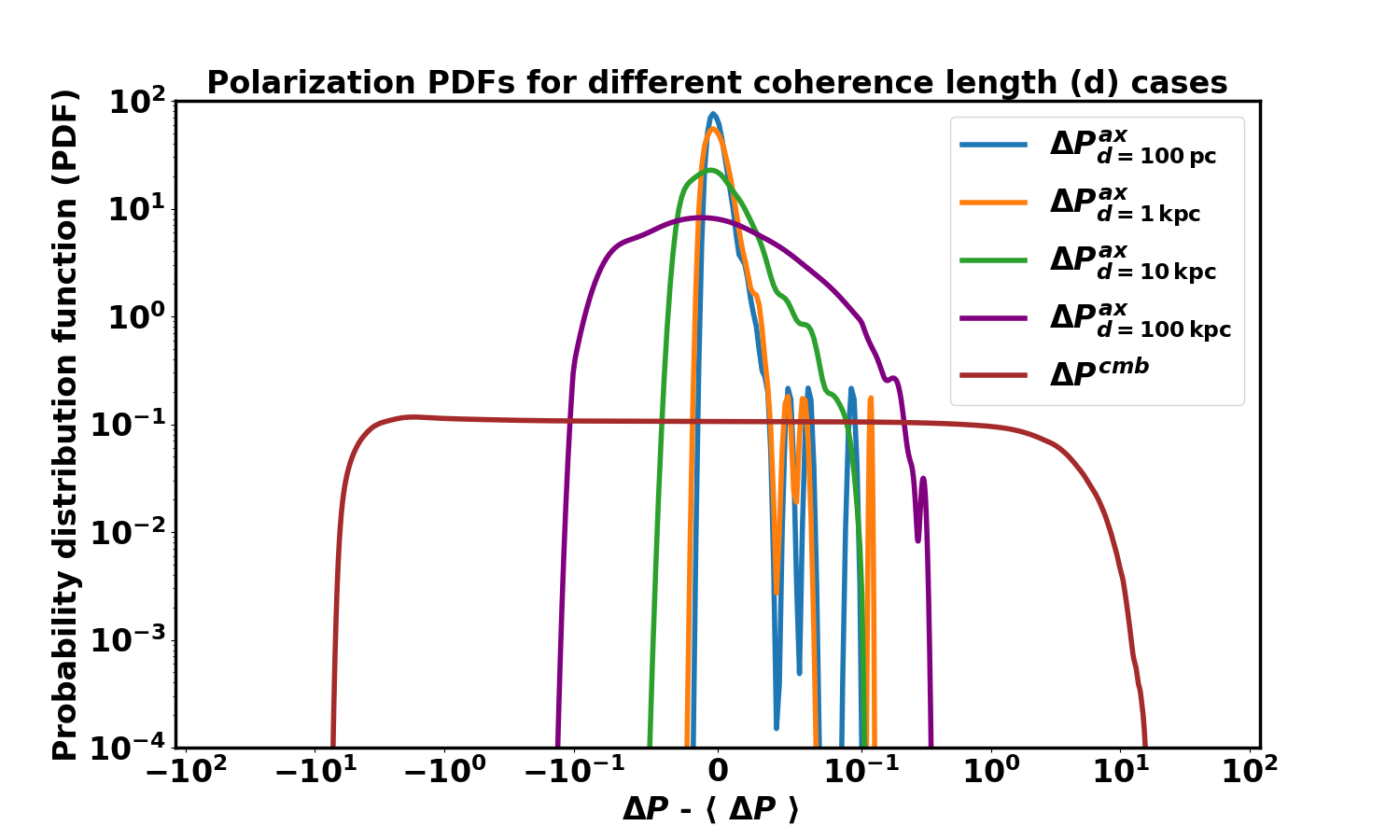}
    \caption{Polarization probability distribution functions (PDFs) for different cases of magnetic field domain sizes. Here $g_{a\gamma} = 10^{-11} \, \mathrm{GeV^{-1}}$.}
       \label{fig:coherpdf}
\end{figure}

The magnetic field coherence length determines the degree of polarization of the ALP distortion signal in CMB. The magnetic field direction determines the polarization of the photons as the conversion to ALPs takes place. If these polarizations are not aligned, depolarization of the signal will take place. In simulations, the magnetic field coherence lengths are taken to be of the order of kiloparsecs to hundreds of kiloparsecs \cite{Marinacci_2018,donnert2018magnetic}. But, that is a highly uncertain quantity depending on the state of the cluster's evolution. The magnetic field coherence length is changed by varying the grid size $d_B$. Thus, the magnetic field direction along any line of sight varies after a distance $d_B$. Also, we have kept the contribution of the longitudinal magnetic field component to the magnetic field magnitude as fixed during this analysis.

We show the results for the coherence scales $d_B = $ 100 kpc, 10 kpc, 1 kpc, and 0.1 kpc for a cluster. The maps and power spectra shown are in the absence of CMB or noise and show the results for the simulated ALP signal only. The temperature signals of the simulated ALP distortion are shown in log intensity values, although the signal will show a negative effect in the CMB temperature maps. 
  The temperature signal remains independent of the magnetic field coherence scale as it depends only on the magnitude of the magnetic field, independent of its direction (see Fig. \ref{fig:coherT}). But the polarization signal exhibits a decrease in strong fluctuations as the coherence length decreases, as shown in Fig. \ref{fig:coher}. The spatial variation of the signal changes as well, depending on the relative directions of the various domains along a line of sight.

The effect of the change in domain size can also be seen by considering the temperature (TT) and polarization (PP) power spectra. In Fig. \ref{fig:coher_TT}, we have plotted the power spectra $D_{\ell}^{TT}$ and $D_{\ell}^{PP}$, where $D_{\ell} = \ell (\ell + 1) C_{\ell} / 2\pi$. We plot the power spectra for the multipole values 1000 to 6000 as lower multipole values will be dominated by CMB, and probing higher multipole values require higher beam resolutions. The TT power spectrum remains invariant of the domain size, while the PP spectrum lowers by about two orders of magnitude as the coherence length is changed from 100 kpc to 0.1 kpc. 
This suppression of power is highlighted in the TP cross spectrum as well, as shown in Fig. \ref{fig:coher_TP}, which shows suppression of power for low coherence scales.
The ALP signal will follow a spatial profile that will vary based on the turbulence associated with the profiles. The power at different angular scales will vary based on this spatial profile of the signal. This shows up in the shape of the power spectra and is highly dependent on the variation of the signal at different angular scales. The oscillations are also a result of the masking kernel corresponding to the cluster region on the sky, where the window function is one for all cluster pixels and zero otherwise \cite{Hivon_2002}.

The ALP signal can also be distinguished from the CMB using its statistical behavior. The polarization probability distribution function (PDF) for different cases of domain sizes is shown in Fig. \ref{fig:coherpdf}. The CMB follows a smooth Gaussian PDF, while the PDF becomes more narrowed and acquires non-Gaussianity as the incoherence increases. This points to the fact that calculating the power spectrum is not enough to analyze the spatial behaviour of the ALP signal, and it requires higher order statistics. The variation of the non-Gaussianity of the signal for different coherence scales can be used to probe the ALP signal using statistics of the polarization signal around the cluster region. If the ALP signal is not present, the statistics will be close to Gaussian for all cluster regions but will vary from cluster to cluster if photon-ALP conversion occurs.

In hydrodynamical simulations, the magnetic field coherence scale is generally taken to be about a few to hundreds of kiloparsecs, with strengths of a few micro-Gauss ($\mu$G) \cite{GOVONI_2004,carilli2002cluster,Ferrari_2008}. This need not be the case, though, especially for non-relaxed clusters, as a higher incoherence length can decrease the power in polarization by two orders of magnitude. {This variation in the signal strength with the coherence scale will also show up in the non-Gaussianity of the signal. The variation in the non-Gaussianity of the polarization observations from cluster regions can be used to probe the ALP coupling for different mass ALPs by measuring the signal as a function of the angular size of the distortion.

\subsection{Inhomogeneity in magnetic field amplitude}
\label{sec:turb_B}

\begin{figure}[h!]
     \centering
\includegraphics[height=6.5cm,width=16cm]{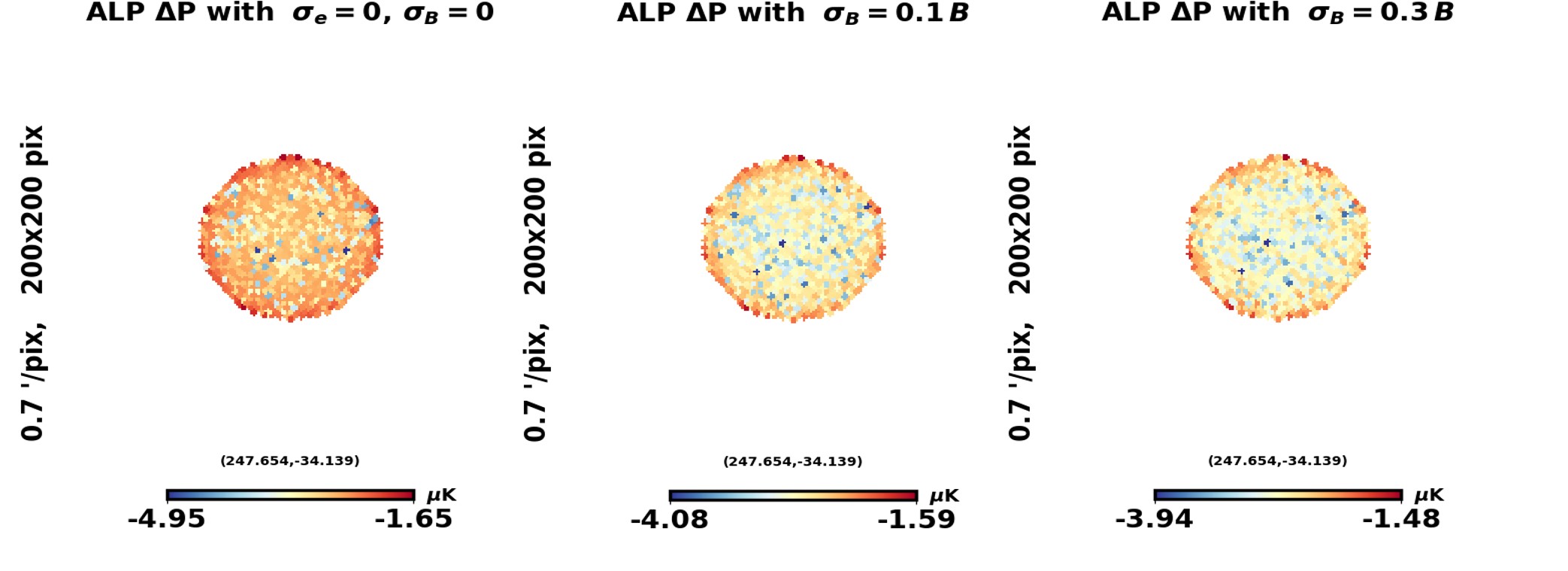}
    \caption{Variation in ALP polarization signal with change in magnetic field strength inhomogeneity. The spatial features do not see a significant change, but the signal magnitude changes. The values are in log scale.}
       \label{fig:Bturb}
\end{figure}

\begin{figure}[h!]
     \centering
\includegraphics[height=7cm,width=11cm]{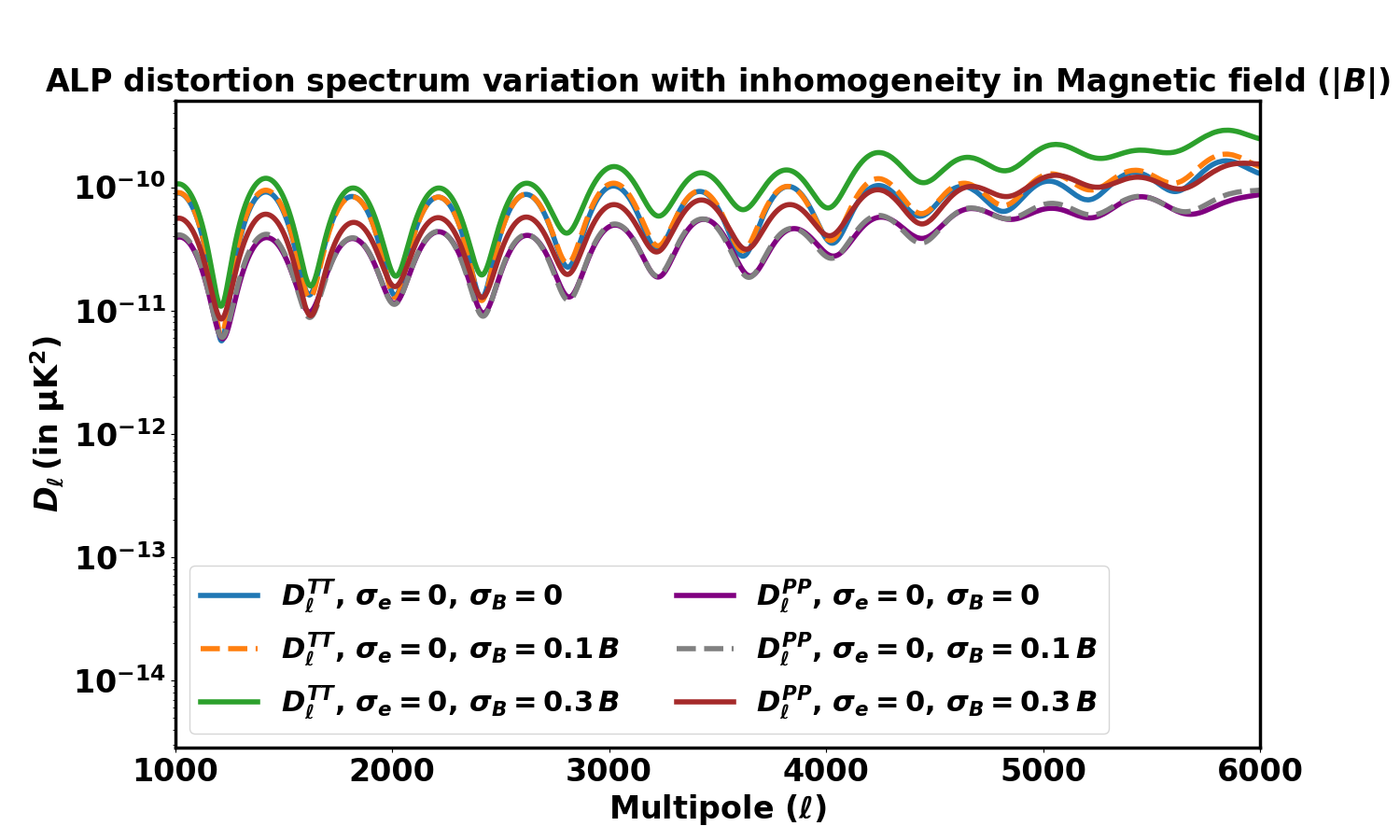}
    \caption{Variation in ALP TT and PP spectra with change in magnetic field strength inhomogeneity. For a high inhomogeneity (standard deviation being 30\% of the mean), the power increases at all multipoles.}
       \label{fig:Bturb_TT}
\end{figure}

\begin{figure}[h!]
     \centering
\includegraphics[height=7cm,width=11cm]{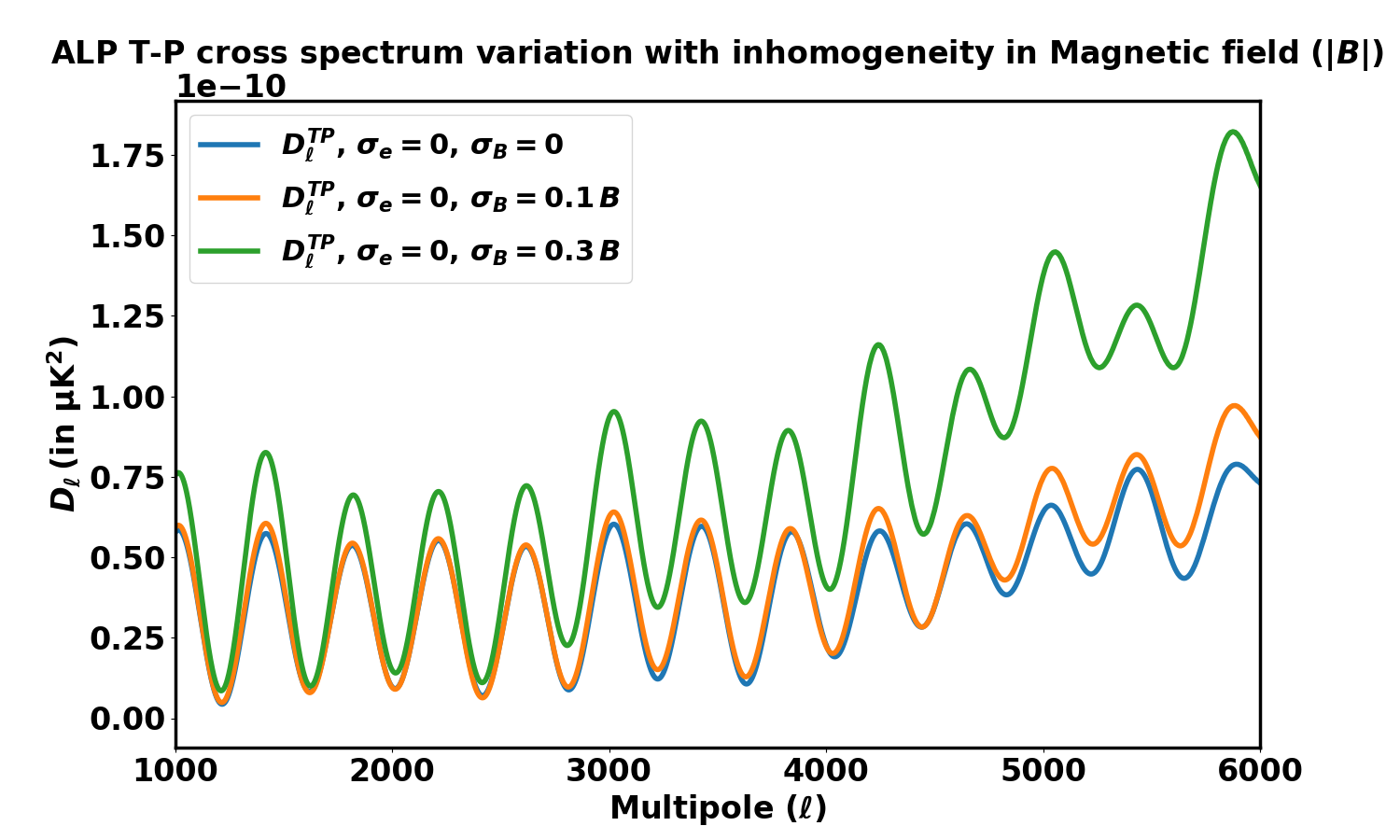}
    \caption{Variation in ALP TP cross spectrum with change in magnetic field strength inhomogeneity. The correlation is higher for the case of high inhomogeneity (standard deviation being 30\% of the mean).}
       \label{fig:Bturb_TP}
\end{figure}

\begin{figure}[h!]
     \centering
\includegraphics[height=7cm,width=11cm]{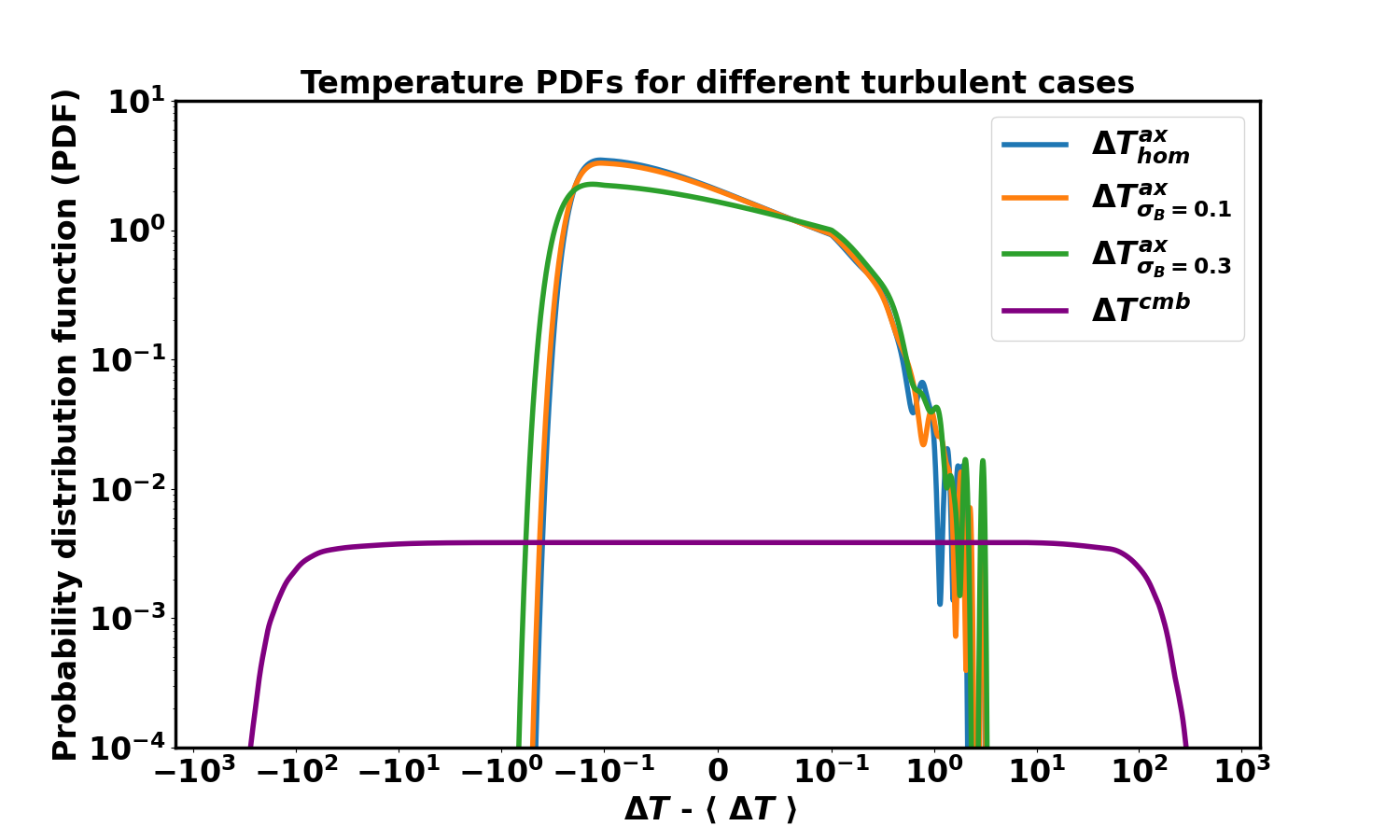}
    \caption{Temperature probability distribution functions (PDFs) for different cases of strength inhomogeneity in the magnetic field. Here $g_{a\gamma} = 10^{-11} \, \mathrm{GeV^{-1}}$.}
       \label{fig:T_B_pdf}
\end{figure}

\begin{figure}[h!]
     \centering
\includegraphics[height=7cm,width=11cm]{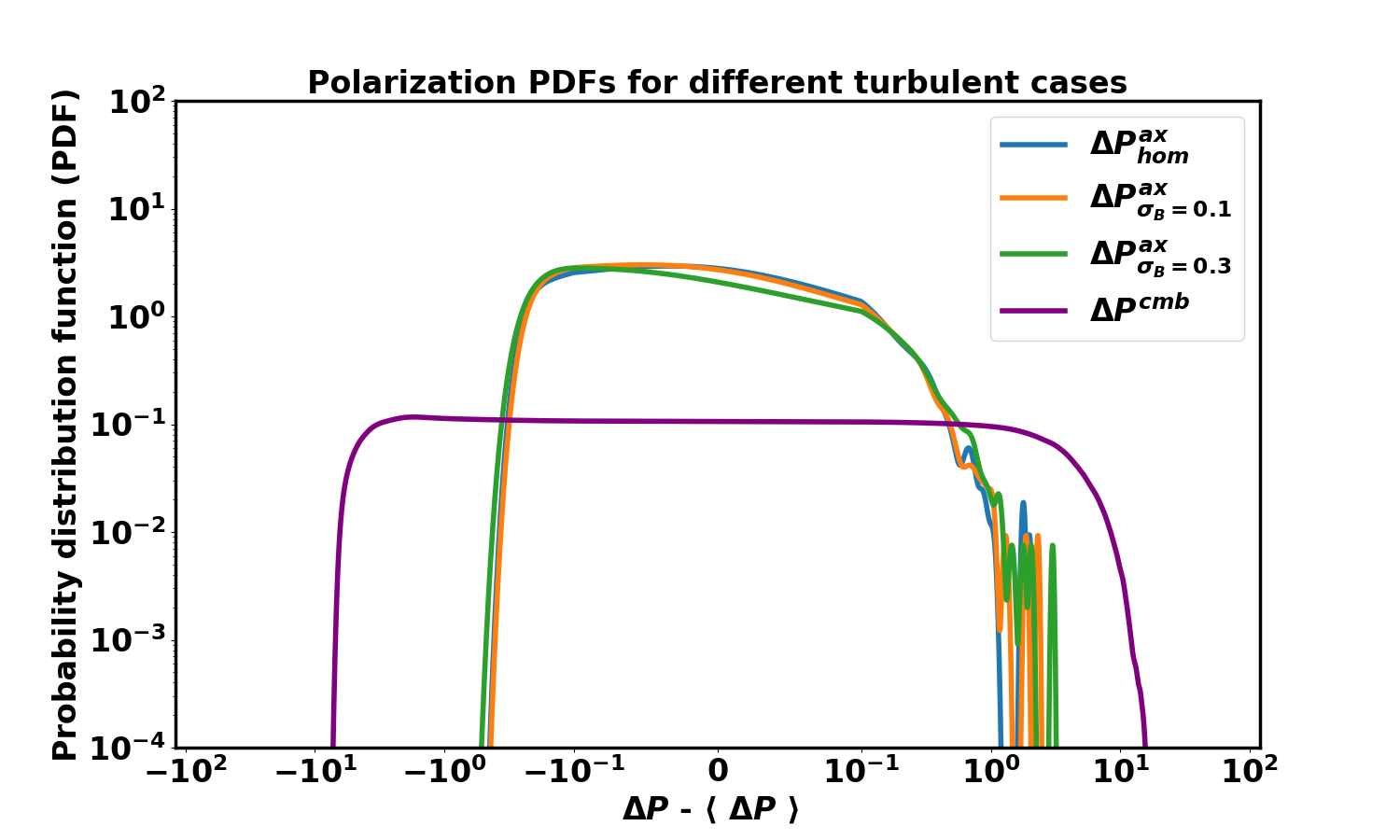}
    \caption{Polarization probability distribution functions (PDFs) for different cases of strength inhomogeneity in the magnetic field. Here $g_{a\gamma} = 10^{-11} \, \mathrm{GeV^{-1}}$.}
       \label{fig:P_B_pdf}
\end{figure}
The change in the transverse magnetic field magnitude affects the strength of the ALP signal via the adiabaticity parameter ($\gamma_{\mathrm{ad}} \propto B_{t}^2$). The presence of inhomogeneity along a single line of sight for multiple conversion locations may or may not lead to an increase in the ALP temperature or polarization signal strength. The fluctuations in the signal generally increase, though, in the presence of inhomogeneity.  In Fig. \ref{fig:Bturb}, we inject inhomogeneity in the magnetic field magnitude at all locations. The inhomogeneity percentage (standard deviation being the reported percentage of the mean) is increased from 0\% (homogeneous) to 10\% to 30\% to check the variation in the ALP signal. The spatial characteristics of the signal over the cluster region do not vary significantly, but the relative strength of the signals along different lines of sight does change. 

The TT and PP power spectra change slightly as the inhomogeneity increases, as shown in Fig. \ref{fig:Bturb_TT}. For a low inhomogeneity (10\%), the power may not necessarily increase at all multipoles, but for a high inhomogeneity, there is an increase in power at all multipoles. Also, the 
TP cross power spectrum shows an increase for the case of high inhomogeneity, as the signal strength increases in both temperature and polarization as shown in Fig. \ref{fig:Bturb_TP}. This happens because the injection of inhomogeneity results in a decrease in magnetic field amplitude in some regions, while an increase in other regions. Thus, over the angular region of a cluster, the fluctuations increase, resulting in an enhancement in the power spectrum. 
The change in strength inhomogeneity in the magnetic field changes the temperature and probability distribution functions as shown in Fig. \ref{fig:T_B_pdf} and Fig. \ref{fig:P_B_pdf}, respectively. This does not produce much of a change in the non-Gaussianity of the ALP signal from the homogeneous case.

\subsection{Turbulent  magnetic field}
\label{sec:turb_Ball}

\begin{figure}[h!]
     \centering
\includegraphics[height=6cm,width=12cm]{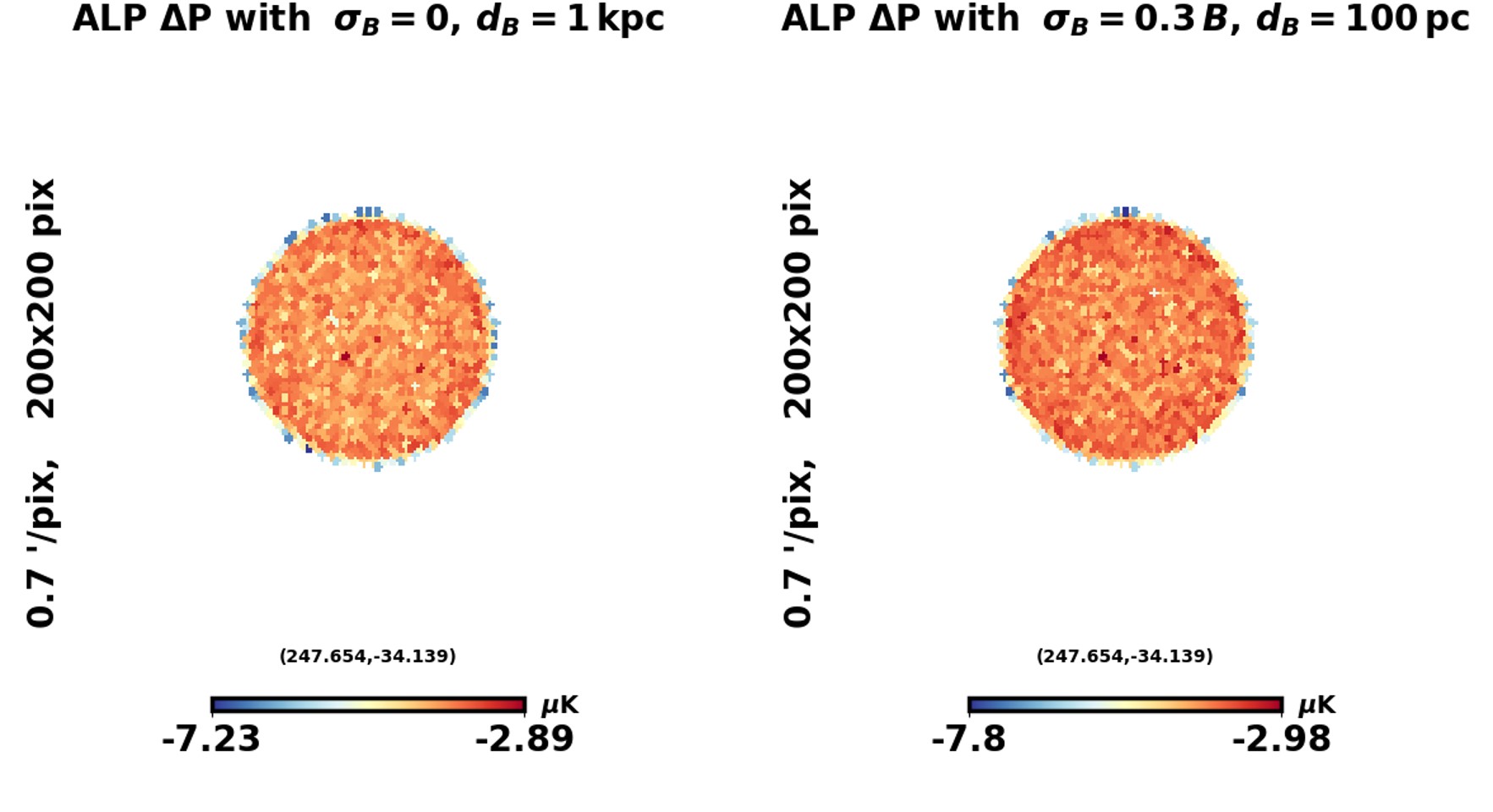}
    \caption{Variation in ALP polarization signal with change in magnetic field strength inhomogeneity (0 to 30\%), as well as the domain size (1 kpc to 100 pc). There is a combined effect on the ALP signal, with the spatial features, as well as the magnitude of the polarization signal changing significantly. The values are in log scale and here $\sigma_e = 0.1 n_e$.}
       \label{fig:B_turb_all_P}
\end{figure}

\begin{figure}[h!]
     \centering
\includegraphics[height=6cm,width=12cm]{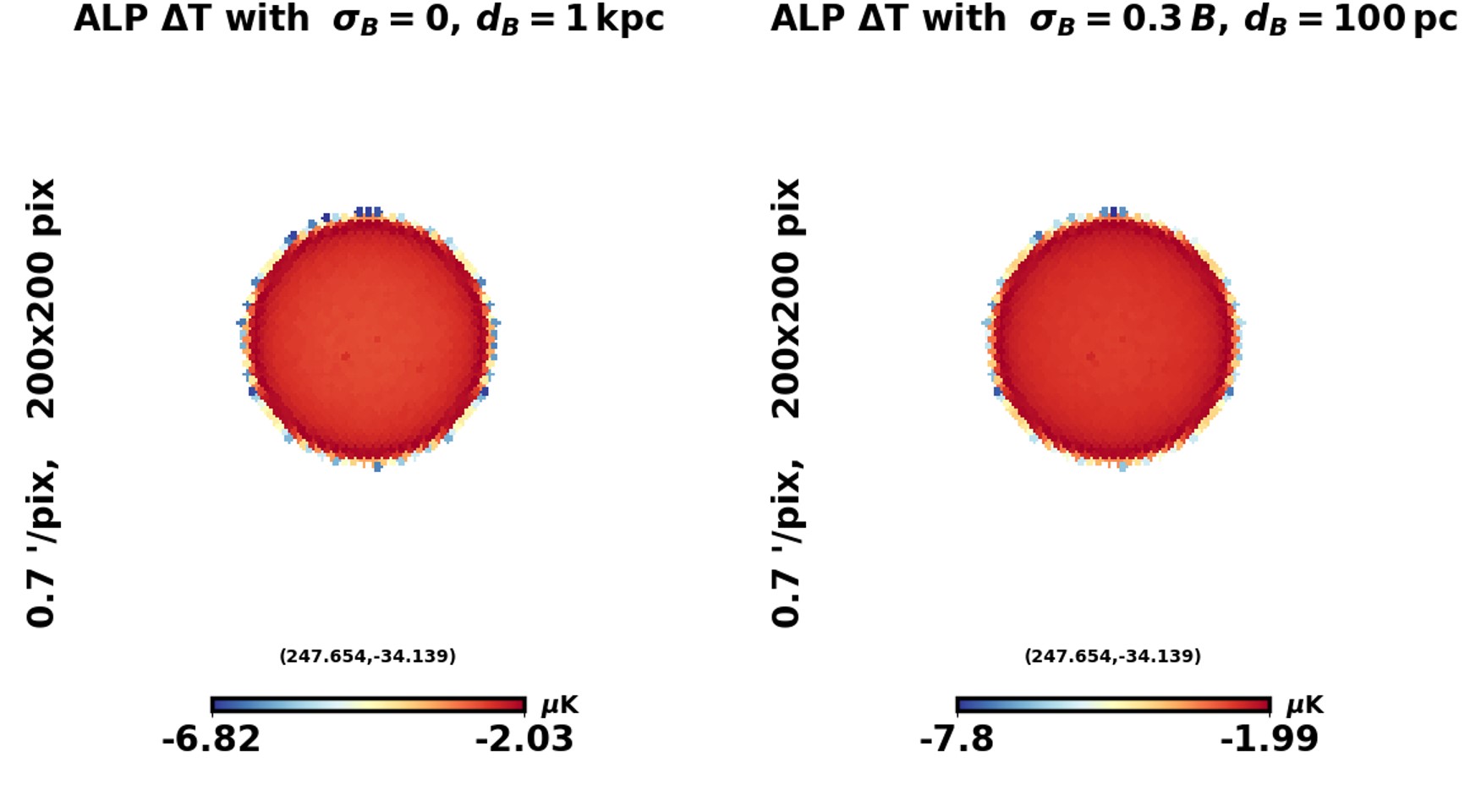}
    \caption{Variation in ALP temperature signal with change in magnetic field strength inhomogeneity (0 to 30\%), as well as the domain size (1 kpc to 100 pc). There is only a change in the magnitude of the signal, which is a result of the turbulence in the magnetic field. The values are in log scale and here $\sigma_e(r) = 0.1 n_{e}(r)$.}
       \label{fig:B_turb_all_T}
\end{figure}

\begin{figure}[h!]
     \centering
\includegraphics[height=7cm,width=11cm]{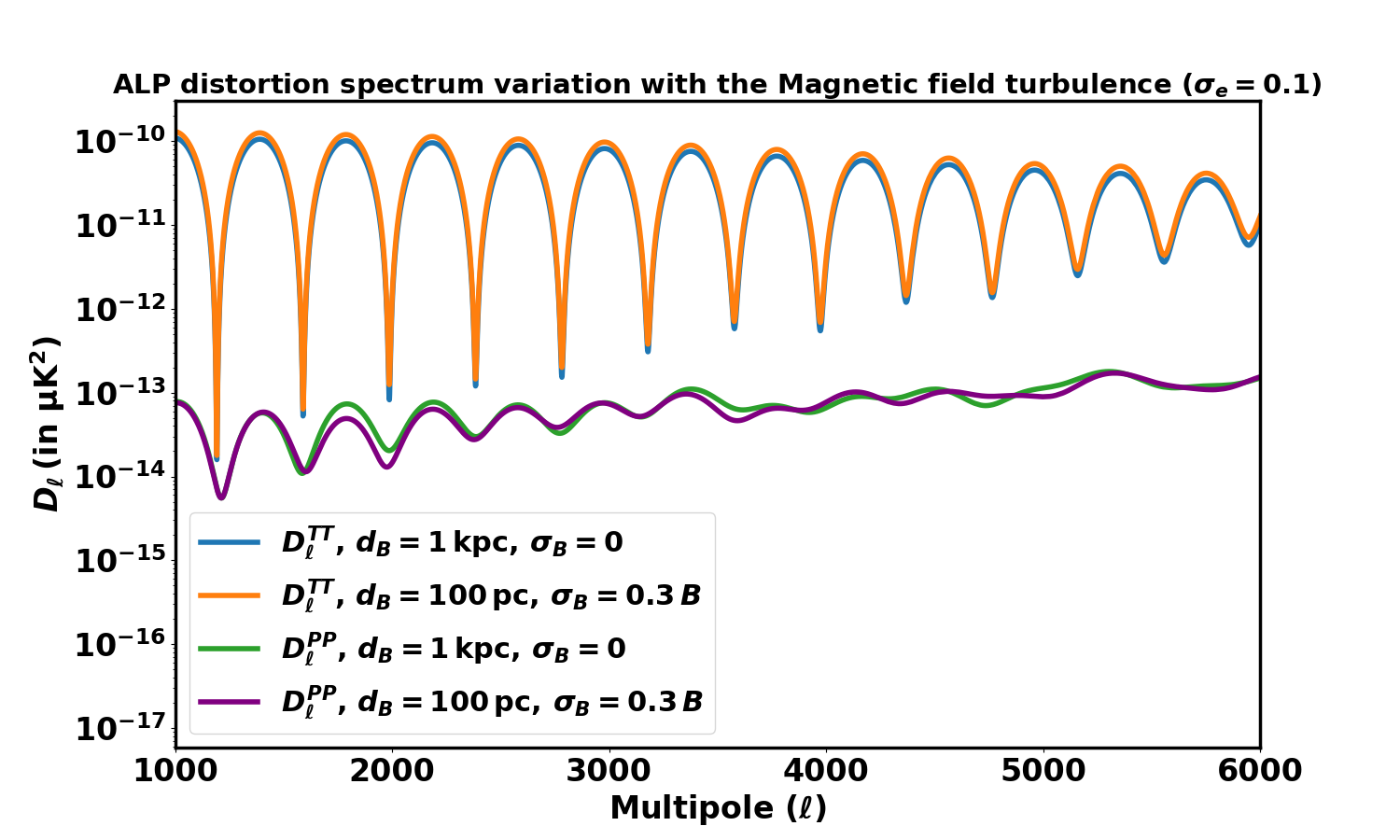}
    \caption{Variation in ALP TT and PP power spectra with change in magnetic field turbulence.}
       \label{fig:Ball_TT}
\end{figure}

\begin{figure}[h!]
     \centering
\includegraphics[height=7cm,width=11cm]{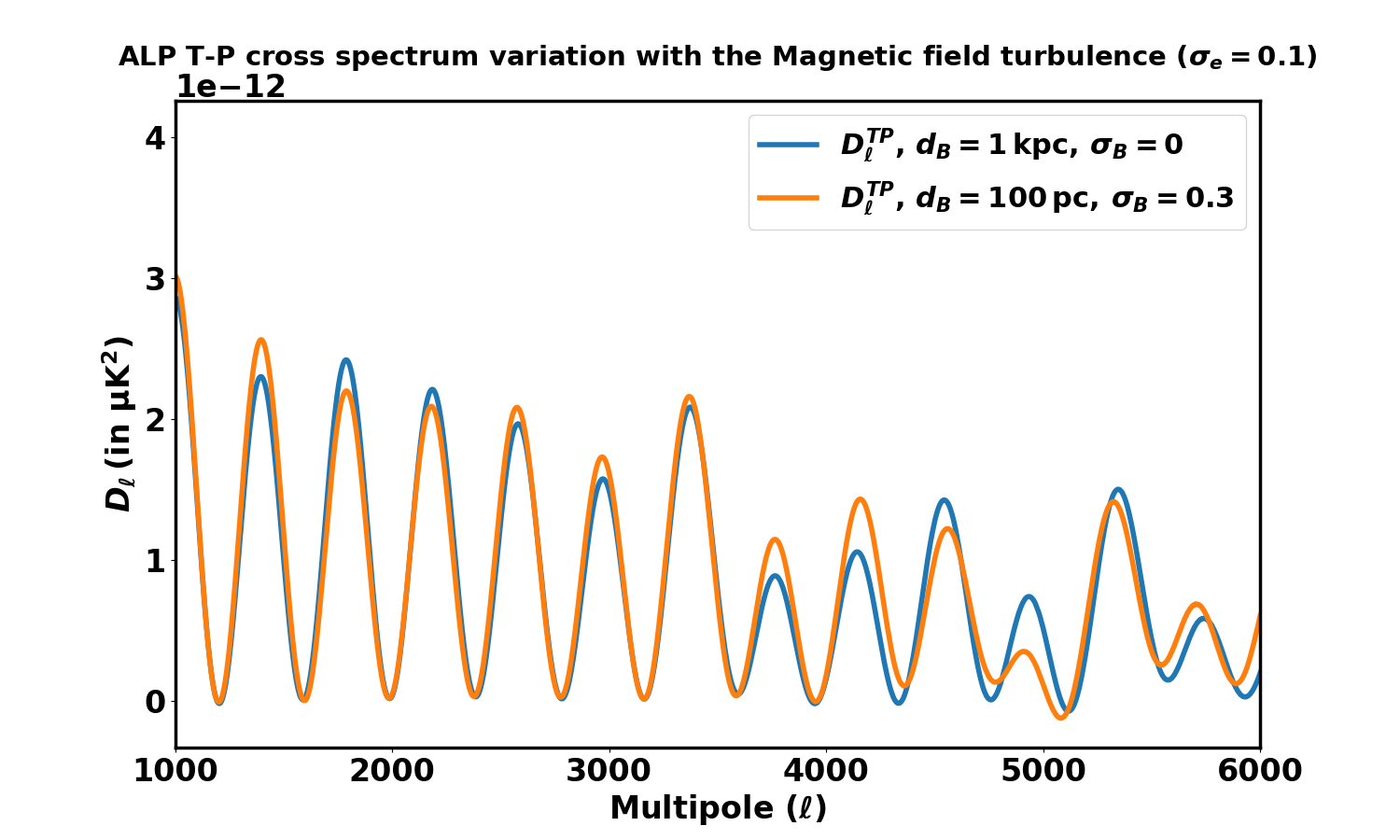}
    \caption{Variation in ALP TP cross spectrum with change in magnetic field turbulence.}
       \label{fig:Ball_TP}
\end{figure}

\begin{figure}[h!]
     \centering
\includegraphics[height=7cm,width=11cm]{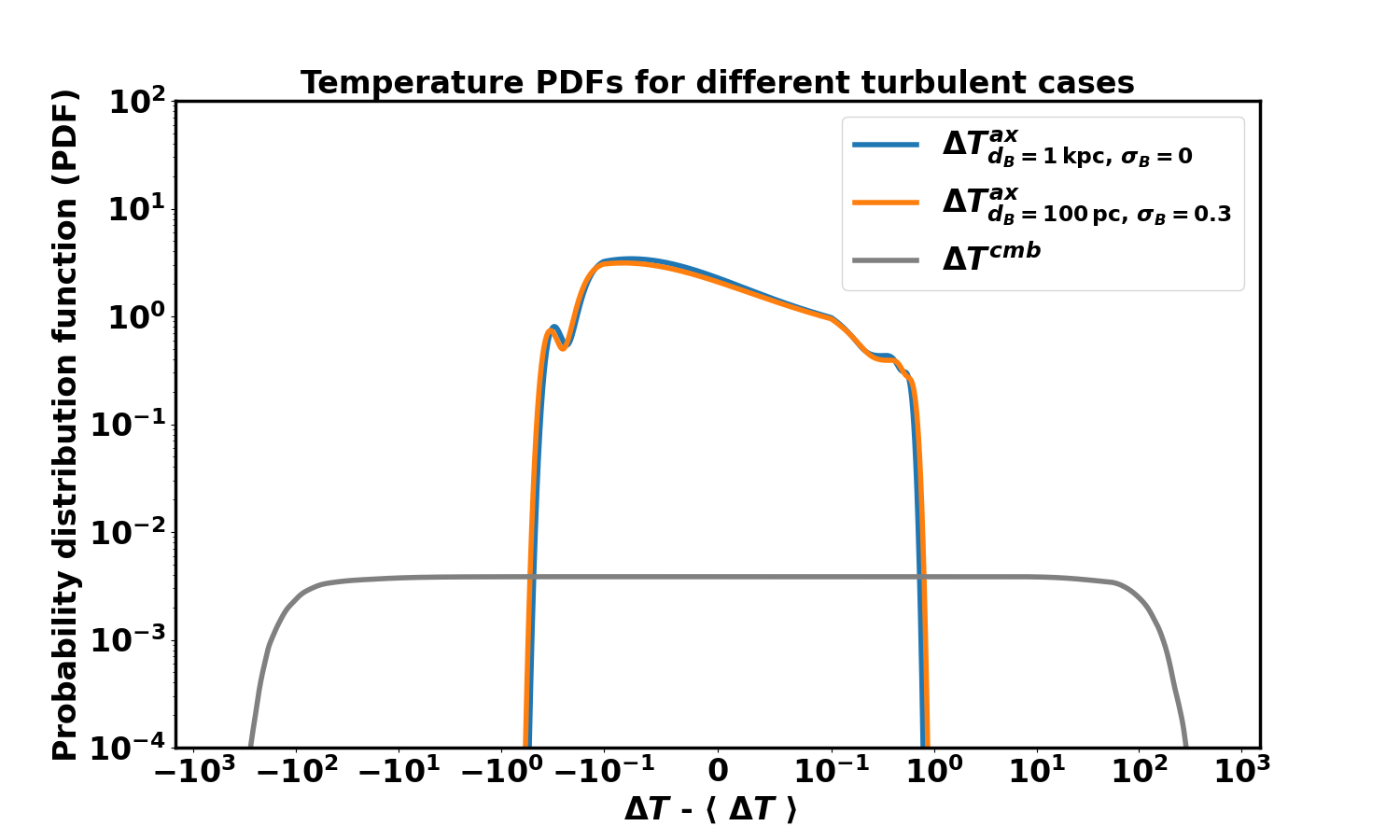}
    \caption{Temperature probability distribution functions (PDFs) for different cases of turbulence in magnetic field. Here $g_{a\gamma} = 10^{-11} \, \mathrm{GeV^{-1}}$.}
       \label{fig:T_Ball_pdf}
\end{figure}
\begin{figure}[h!]
     \centering
\includegraphics[height=7cm,width=11cm]{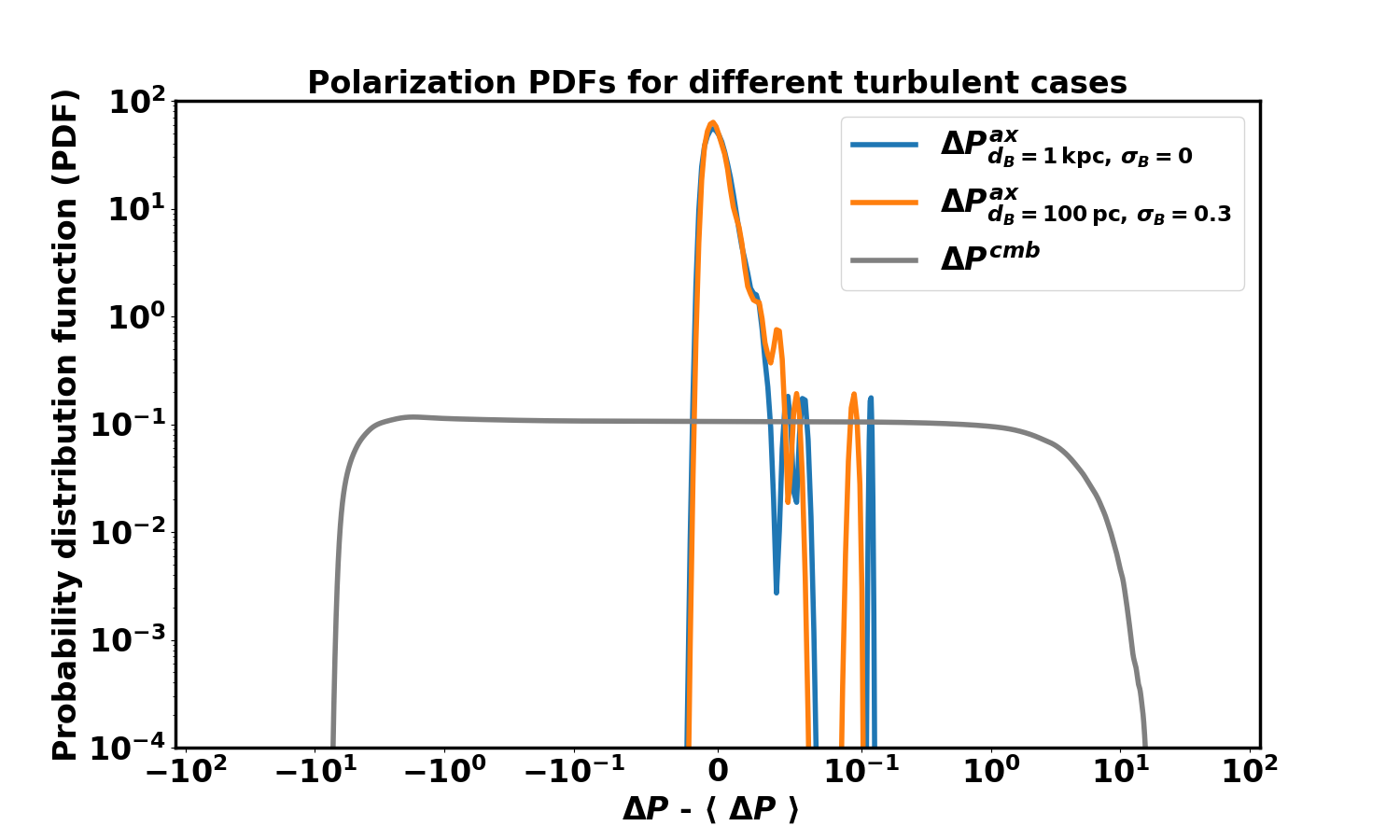}
    \caption{Polarization probability distribution functions (PDFs) for different cases of turbulence in magnetic field. Here $g_{a\gamma} = 10^{-11} \, \mathrm{GeV^{-1}}$.}
       \label{fig:P_Ball_pdf}
\end{figure}
A change in both the coherence length and strength inhomogeneity in the magnetic field will change the ALP polarization signal in a combined effect. For a lower coherence length, the signal gets suppressed in polarization, while the strength inhomogeneity 
 in the magnetic field may enhance it, as shown in Fig. \ref{fig:B_turb_all_P}.  A change in coherence scale from 1 kpc (left) to 100 pc (right) reduces the polarization signal way more than the enhancement from an injection of 30\% inhomogeneity (standard deviation being 30\% of the mean). The change in ALP temperature signal will only be an effect of the strength inhomogeneity in the magnetic field, as it is independent of the coherence scale as shown in Fig. \ref{fig:B_turb_all_T}.

The TT power spectrum increase, as shown in Fig. \ref{fig:Ball_TT} is only due to magnetic field strength inhomogeneity. The PP spectrum is impacted more by the reduction in coherence length. {Also, the TP cross spectrum for the two cases (see Fig. \ref{fig:Ball_TP}) shows disparities mostly at higher multipole values, as it is mostly the smaller angular scales that are affected by turbulence. The PDFs for the cases are shown in Fig. \ref{fig:T_Ball_pdf} (for temperature) and Fig. \ref{fig:P_Ball_pdf} (for polarization), where there is a slight change in PDF for the temperature case, but a greater change in polarization, as the polarization is also affected by the change in coherence length. }
{
The non-Gaussianity of the signal in polarization will vary more though, as shown in Fig. \ref{fig:P_Ball_pdf}, can be used to probe the ALP signal from galaxy clusters, as the statistics will vary from one cluster region to another depending on the turbulence in those regions, as well as the ALP coupling with photon.} 
\subsection{Turbulent electron density}
\label{sec:turb_ne}

\begin{figure}[h!]
     \centering
\includegraphics[height=6.5cm,width=15cm]{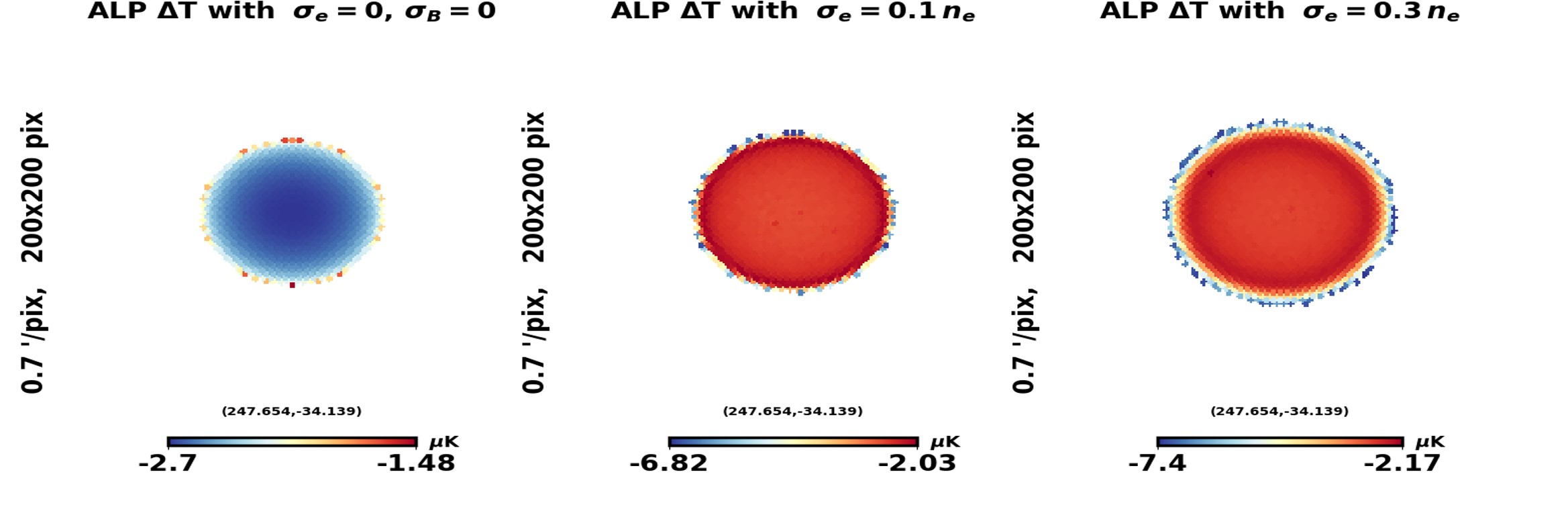}
    \caption{Variation in ALP temperature signal with change in electron density turbulence. The ALP distortion suffers a change in shape, extent, as well as strength, with the spatial features getting significantly changed for the high turbulence (30\%) case. The values are in log scale.}
       \label{fig:neturb}
\end{figure}

\begin{figure}[h!]
     \centering
\includegraphics[height=7cm,width=11cm]{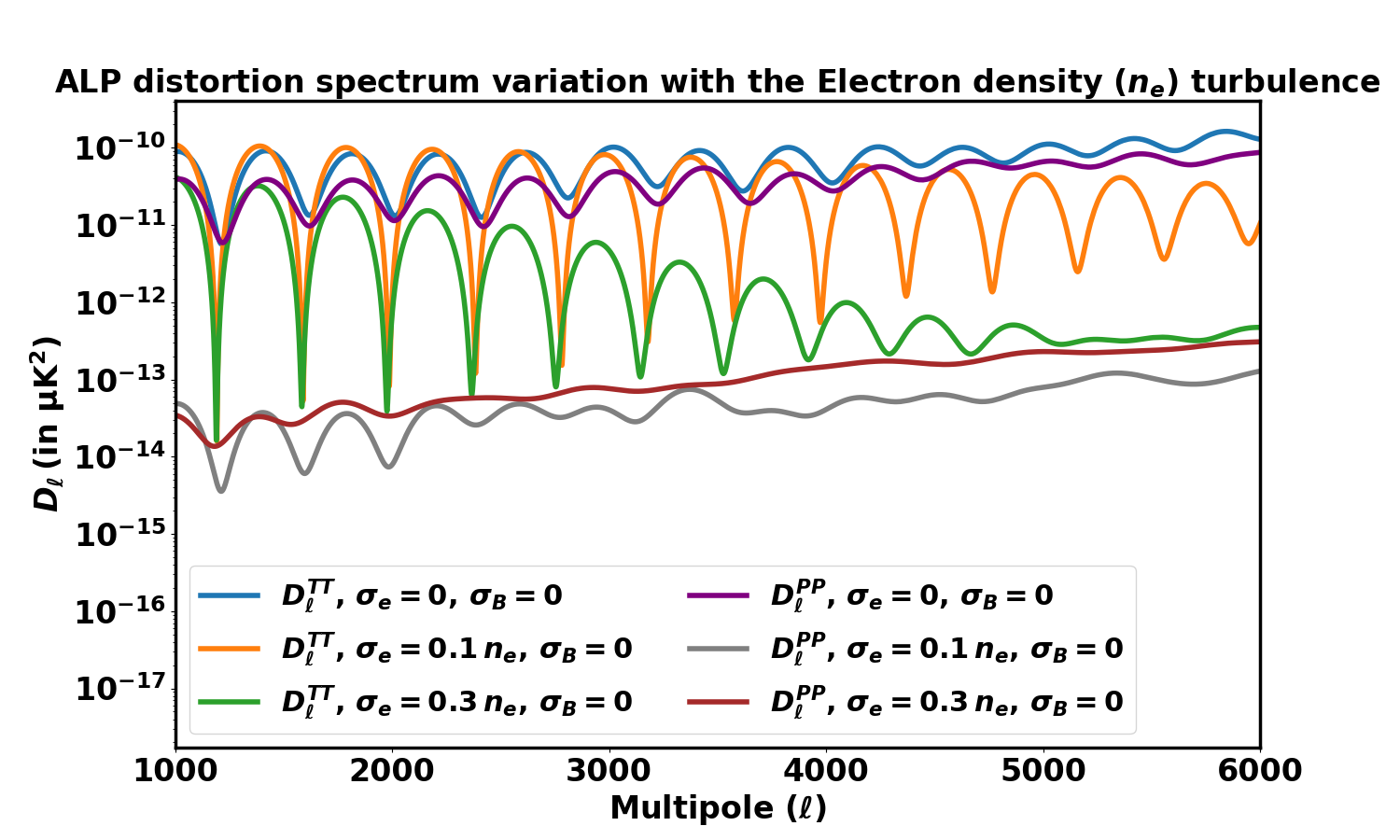}
    \caption{Variation in ALP TT and PP power spectra with change in electron density turbulence. The PP spectrum increases and approaches the TT spectrum at high multipoles for all cases.}
       \label{fig:neturb_TT}
\end{figure}

\begin{figure}[h!]
     \centering
\includegraphics[height=7cm,width=11cm]{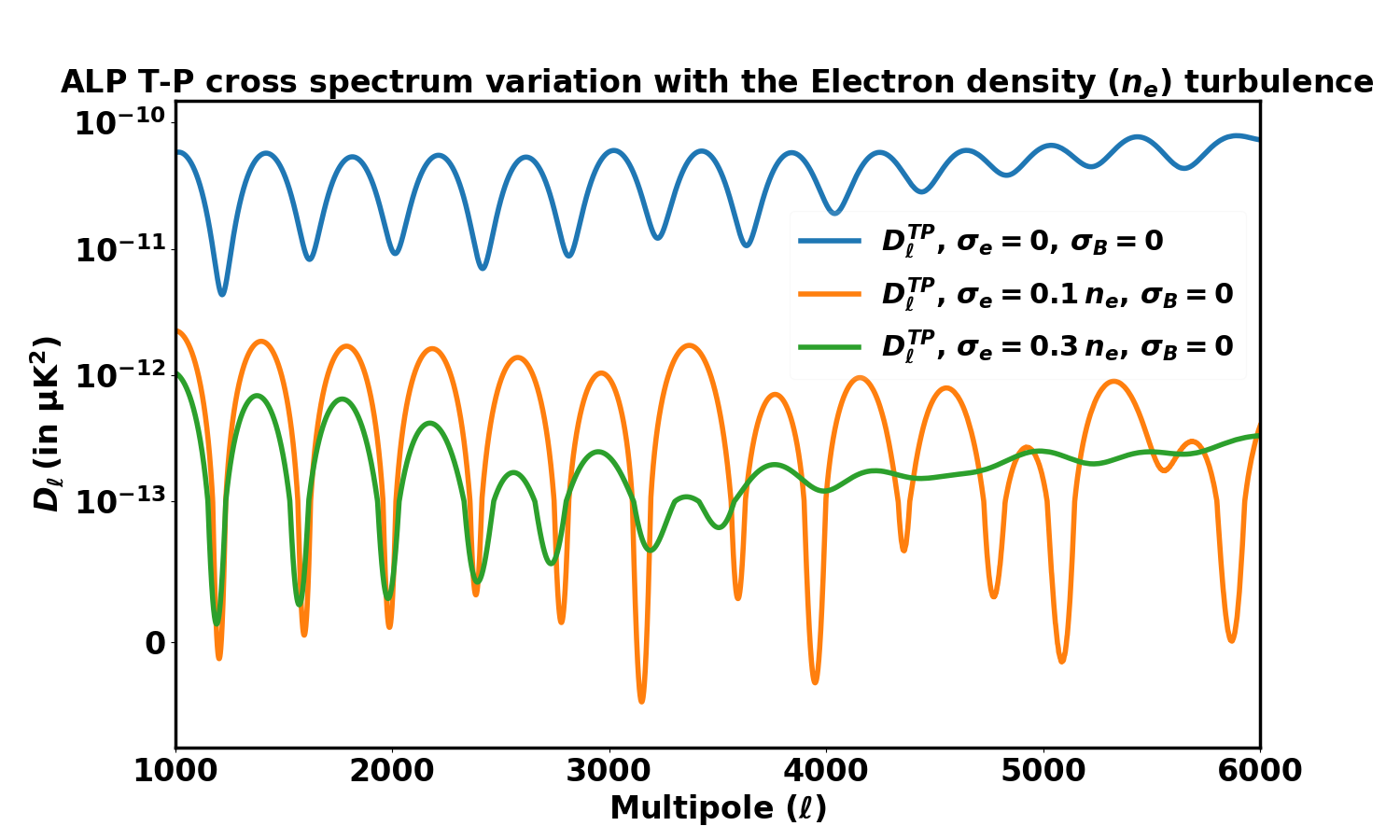}
    \caption{Variation in ALP TP cross spectrum with change in electron density turbulence. The injection of turbulence in electron density leads to a significant change in the TP spectrum.}
       \label{fig:neturb_TP}
\end{figure}

\begin{figure}[h!]
     \centering
\includegraphics[height=7cm,width=11cm]{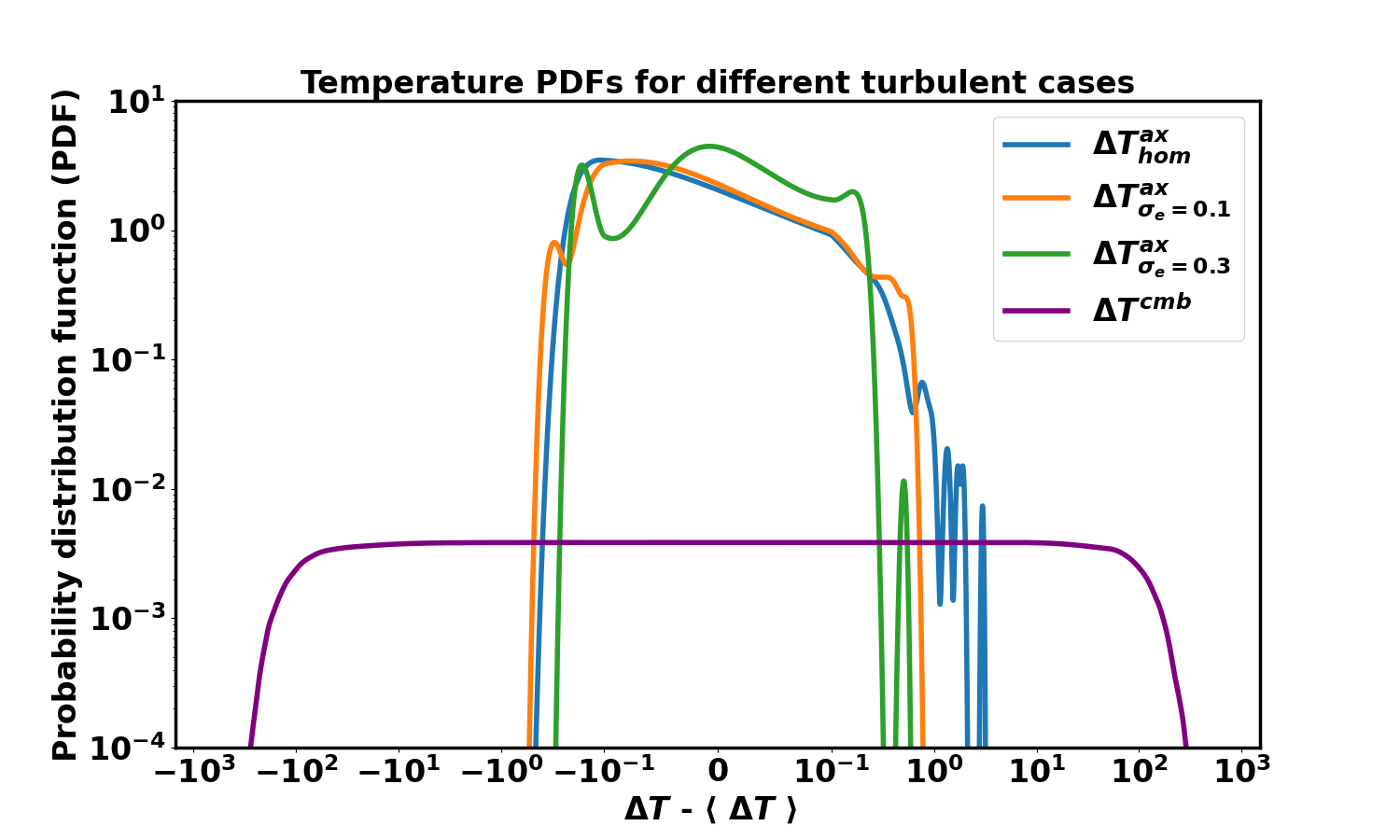}
    \caption{Temperature probability distribution functions (PDFs) for different cases of turbulence in electron density. Here $g_{a\gamma} = 10^{-11} \, \mathrm{GeV^{-1}}$.}
       \label{fig:T_e_pdf}
\end{figure}
A change in electron density produces a much greater change in the ALP signal for the same percentage of inhomogeneity in the magnetic field amplitude. 
In the absence of turbulence in electron density, for a smooth radially varying profile, the ALP signal is generated in a spherical shell within the cluster, which is projected on the two-dimensional sky as a circular disk. The size of the disk depends on the mass of ALPs, with a larger-sized disk for low-mass ALPs, and a smaller one for high-mass ALPs, as the electron density decreases in the outer regions of the cluster \cite{Mukherjee_2020,Mehta:2024wfo,Mehta:2024sye}. Also, the signal generally shows an increase in strength in the outer regions of the disk before sharply falling off, as shown in Fig. \ref{fig:neturb}. 

In the presence of turbulence, the resonant condition ($m_a^2 = m_{\gamma}^2 \propto n_e$) is not limited to a spherical shell within the cluster. This effect for the ALP temperature signal is shown in Fig. \ref{fig:neturb}. We compare the case of no turbulence with the cases of 10\% and 30\% turbulence in electron density (standard deviation being the reported percentage of the mean). The resonant locations depend on the local turbulent effects and distort the circular disk, which is much greater when the turbulence is high (30\%). The resonant condition is satisfied for some outer regions in the cluster as well, thus increasing the size of the signal disk projected on the sky. This change is observed in both temperature and polarization signals and the extent of the disk increases with an increase in turbulence. The turbulent electron density also determines the strength of the signal via the electron density gradient that changes the probability of conversion ($\gamma_{\mathrm{ad}} \propto |\nabla n_e|^{-1}$). 
Thus, the presence of turbulence in electron density changes the resonant locations and also the strength of conversion that depends on the local gradient at the new locations. The number of resonances for a particular mass ALP will generally increase with an increase in turbulence, but the overall strength of the signal decreases as electron density fluctuations increase, thus increasing its gradient and, hence, lowering the conversion probabilities for the resonant locations.

In terms of the power spectrum of the signal, the cases of turbulence and no turbulence show a large difference between the temperature and polarization powers. In the case of no turbulence, the number of resonant locations along any given line of sight is restricted to two, thus, the depolarization is low even with a random orientation of the magnetic field. That is not the case for turbulent electron densities, as multiple resonant locations with lower conversion probabilities may lead to depolarization of the signal (see Fig. \ref{fig:neturb_TT}). The PP power spectrum starts approaching the TT power spectrum for high multipoles as at small angular scales, it is the individual bright line of sights that dominate the power.  
The TP cross spectrum shows a significant decrease in strength due to the effect of turbulence, as shown in Fig. \ref{fig:neturb_TP}. This happens as both the temperature and polarization signals decrease with an increase in turbulence, owing to increased electron density gradients.

A change in the electron density turbulence produces a significant change in the PDF of the signal, as the PDFs become narrower and the shape  also changes, with an increase in turbulence. This is shown in Fig. \ref{fig:T_e_pdf}. This highlights a significant change in the non-Gaussian statistics of the signal with an increase in turbulence in electron density, as compared to the magnetic field strength inhomogeneity. The non-Gaussianity of the signal will vary both in temperature and polarization in case of turbulence in electron density. The study of the distribution of the distortion in temperature and polarization around galaxy clusters can hint towards the existence of the ALP signal.

\subsection{Turbulent electron density and magnetic field}\label{sec:turb_neB}

The injection of strength inhomogeneity in both the electron density and magnetic field changes the spatial shape as well as the strength of the ALP signal, in a combined effect. This can be seen in Fig. \ref{fig:neBturb}, where a 10\% (standard deviation being 10\% of the mean) inhomogeneity in both profiles increases the extent and decreases the mean amplitude of the ALP polarization signal (top $\rightarrow$ center), but with higher fluctuations. The increase in signal disk size occurs as inhomogeneity leads to resonant locations in relatively outer regions of the cluster as well. The change in signal strength is a combined effect of the electron density gradient ($\gamma_{\mathrm{ad}} \propto |\nabla n_e|^{-1}$)  and magnetic field variation ($\gamma_{\mathrm{ad}} \propto B_{t}^2$), but the effect of resonances getting weaker due to more fluctuating electron density dominates. A further change in magnetic field inhomogeneity to 30\% (center $\rightarrow$ right) results in a more fluctuating signal ($\gamma_{\mathrm{ad}} \propto B_{t}^2$). On the other hand, keeping the magnetic field inhomogeneity at 10\% and increasing the electron density turbulence to 30\% drastically changes the spatial features, extent, and the magnitude of the signal (center $\rightarrow$ left). This happens as even more resonant locations are generated with large electron density gradients as a result of inhomogeneities ($\gamma_{\mathrm{ad}} \propto |\nabla n_e|^{-1}$).   A further increase in magnetic field inhomogeneity to 30\% then results in a more fluctuating signal ($\gamma_{\mathrm{ad}} \propto B_{t}^2$) (left $\rightarrow$ bottom). The following schematic can be considered for temperature signal as well, where a change in the magnetic field inhomogeneity will mostly change the magnitude of the signal, while a change in electron density inhomogeneity will change the magnitude as well as the spatial extent of the signal.

At the power spectra level, the TT and PP spectra are mainly affected by the inhomogeneity in electron density as seen in Fig. \ref{fig:neBturb_TT}. The difference between the TT and PP spectra for the case of no inhomogeneity is low because of the number of resonances being restricted to two, which limits the depolarization of the signal. The change in magnetic field inhomogeneity mostly affects only the amplitude of the power spectra ($\gamma_{\mathrm{ad}} \propto B_{t}^2$), with increased power for high inhomogeneity (30\%), as there are more signal fluctuations that follow the fluctuations in the magnetic field. Injection of electron density turbulence significantly reduces the temperature and polarization powers as a result of more but weak resonances and increased depolarization of the signal. Also, it is the fluctuations in electron density that impact the TP power spectrum more as compared to fluctuations in the magnetic field, as a result of depolarization and weak resonances as seen in Fig. \ref{fig:neBturb_TP}. 

\begin{figure}[h!]
     \centering
\includegraphics[height=19cm,width=19cm]{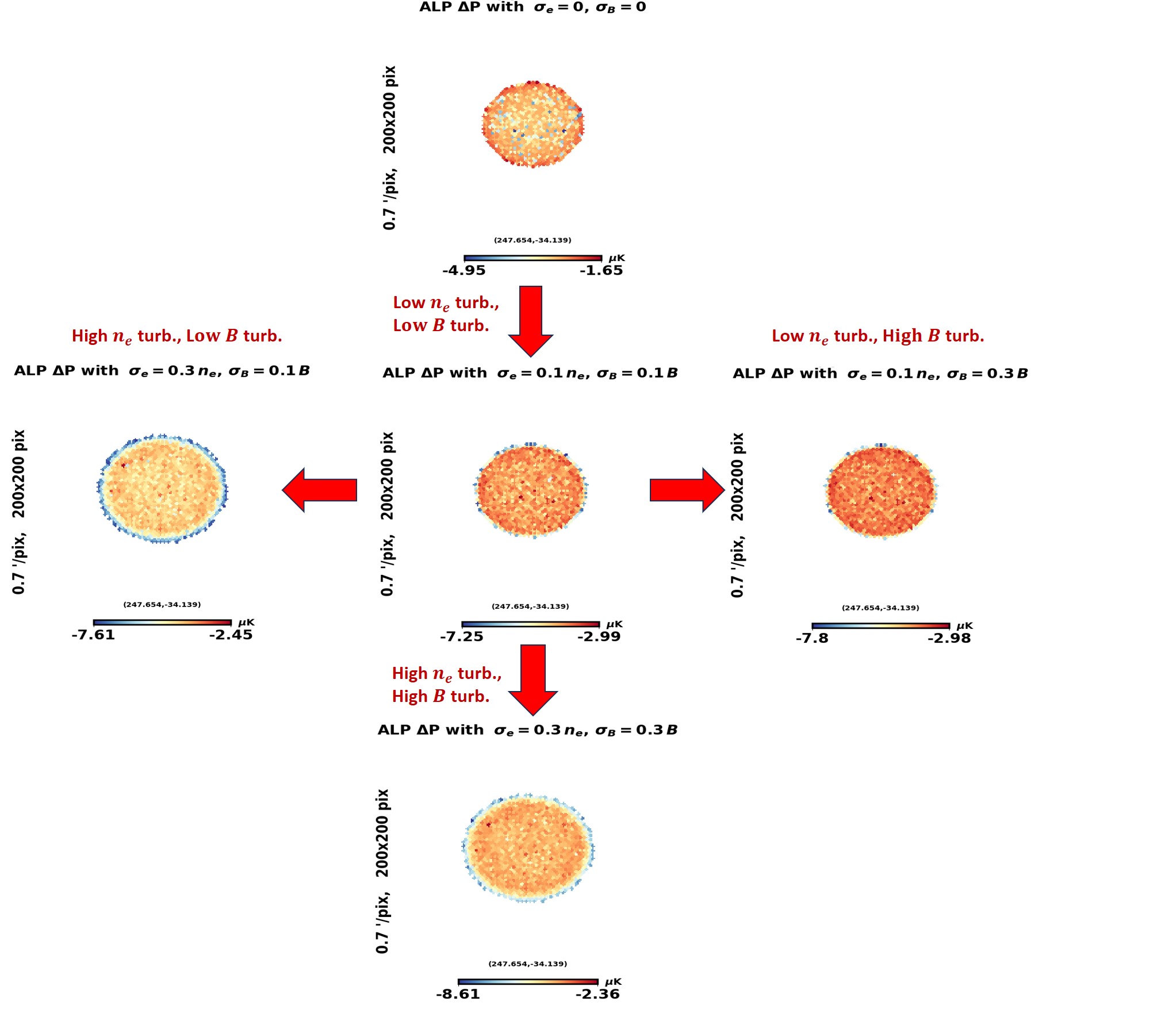}
    \caption{Variation in ALP polarization signal with change in both the electron density and magnetic field strength inhomogeneity. The values are in log scale.}
       \label{fig:neBturb}
\end{figure}

\begin{figure}[h!]
     \centering
\includegraphics[height=7cm,width=11cm]{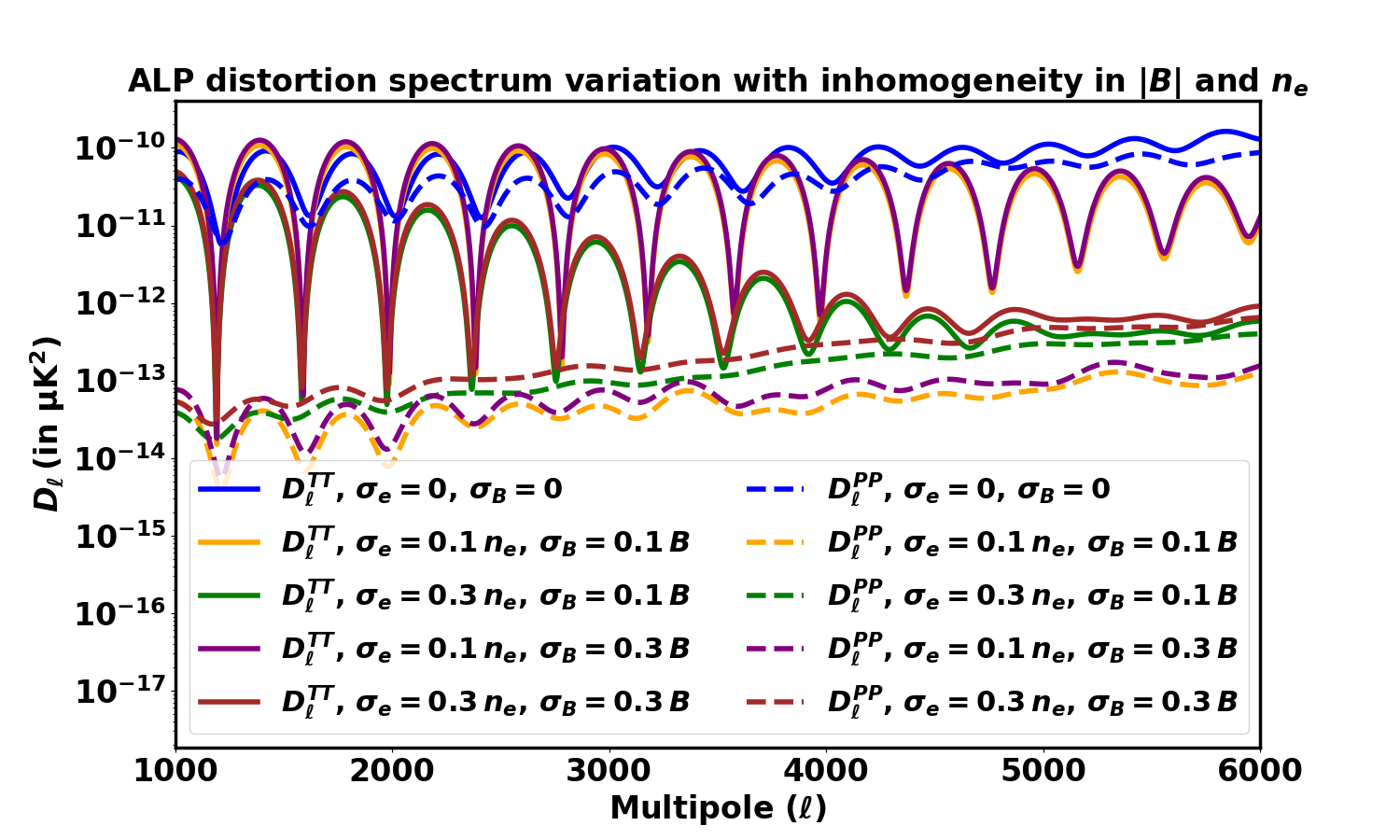}
    \caption{Variation in ALP TT and PP spectra with change in both the electron density and magnetic field strength inhomogeneity. }
       \label{fig:neBturb_TT}
\end{figure}

\begin{figure}[h!]
     \centering
\includegraphics[height=7cm,width=11cm]{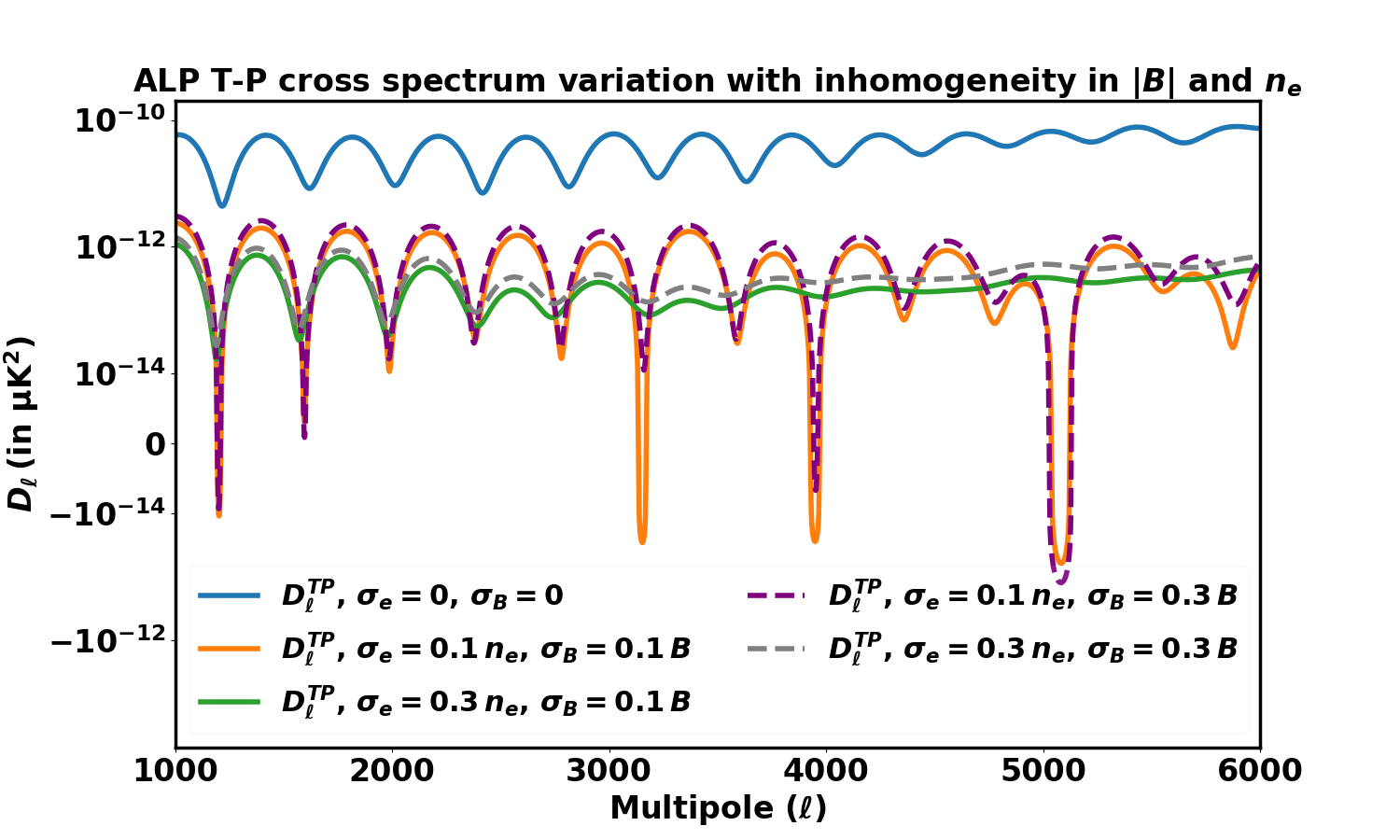}
    \caption{Variation in ALP TP cross spectrum with change in both the electron density and magnetic field strength inhomogeneity.}
       \label{fig:neBturb_TP}
\end{figure}

\begin{figure}[h!]
     \centering
\includegraphics[height=7cm,width=11cm]{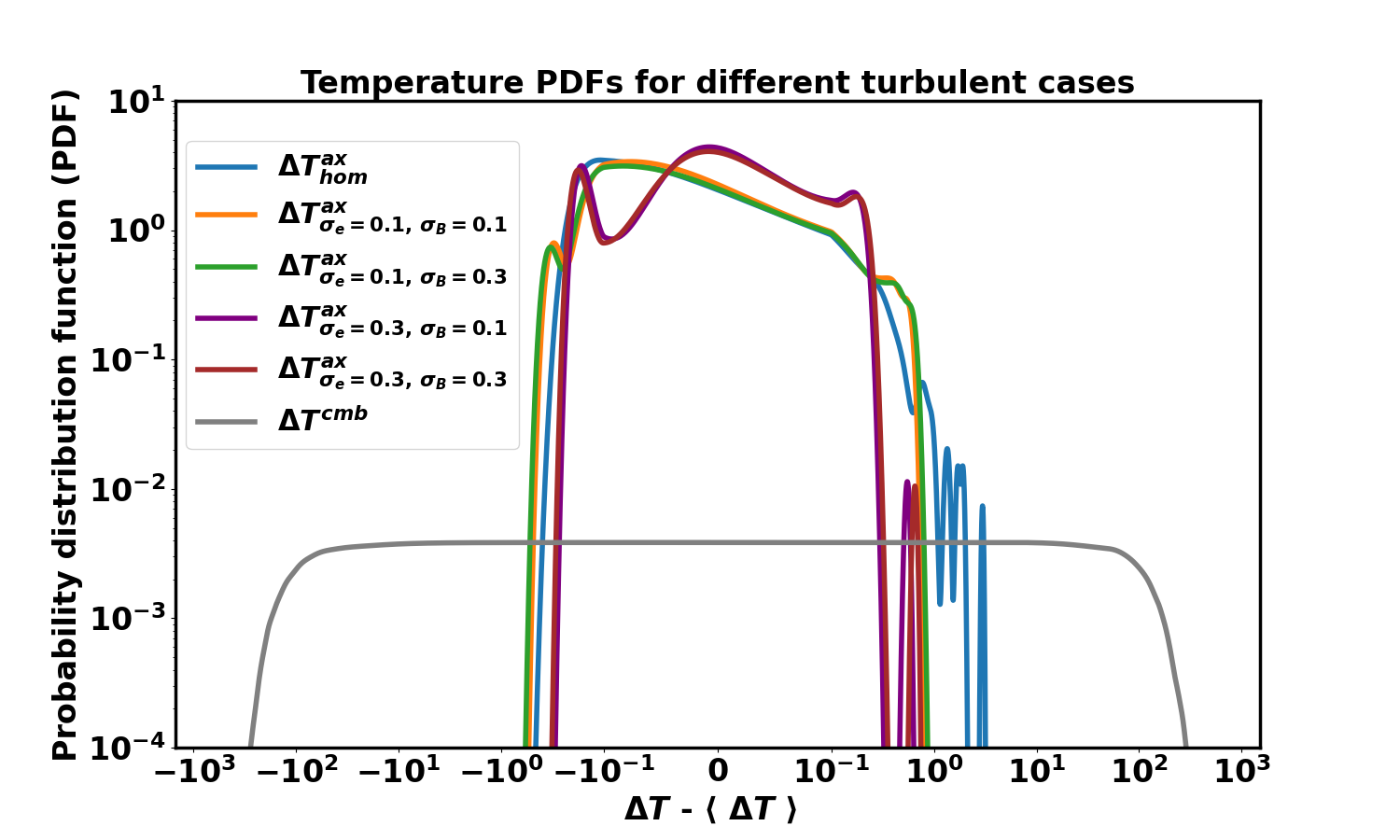}
    \caption{Temperature probability distribution functions (PDFs) for different cases of turbulence in electron density and strength inhomogeneity in the magnetic field. Here $g_{a\gamma} = 10^{-11} \, \mathrm{GeV^{-1}}$.}
       \label{fig:T_eB_pdf}
\end{figure}

{The temperature PDFs for the different cases of turbulence are shown in Fig. \ref{fig:T_eB_pdf}. The PDFs do not show a significant change when the magnetic field strength inhomogeneity is increased, while the shape of the distribution drastically changes as the turbulence in electron density is increased. The non-Gaussianity can be explored in both the temperature and polarization signals.}

\section{Implications for future analyses}
\label{sec:implications}
The ALP signal for the case of homogeneous smooth-radially varying coherent electron density and magnetic field follows a spatial profile that increases in the outer regions and decreases again. But that is an ideal assumption, and there will always be turbulence associated with these profiles, which will be a result of the strength inhomogeneity and incoherence due to energy injection from gravitational and astrophysical phenomena \cite{subramanian2006evolving,schekochihin2006turbulence,xu2009turbulence,ensslin2006magnetic}. These will change the strength, as well as the spatial profile of the ALP distortion signal. The simulation setup developed in this work will provide the means to simulate the ALP signal in a more realistic scenario of the presence of inhomogeneity in cluster profiles, with control on the magnitude of both the electron density and magnetic field profiles and also on the magnetic field direction.

The three-dimensional simulation setup will be integrated into the multi-band framework \texttt{SpectrAx} \cite{Mehta:2024sye} that uses multi-band observations of galaxy clusters from various surveys to obtain their astrophysical information (magnetic field profiles from radio telescopes such as the SKA \cite{Carilli_2004,braun2019anticipated}, electron densities and temperature from X-ray surveys such as eROSITA \cite{predehl2021erosita,bulbul2024srgerosita}, and redshifts from optical surveys such as the {Dark Energy Spectroscopic Instrument (DESI) \cite{aghamousa2016desi}, Rubin Observatory \cite{ivezic2019lsst,abell2009lsst}, etc. } and will be able to place constraints on the photon-ALP coupling constant $g_{a\gamma}$ from both CMB temperature and polarization observations. 
 Also, the case of spectral correlation of the TP cross spectrum for different frequencies, as well as the non-Gaussianity of the ALP signal, will serve as an independent search for ALPs as shown in Sec. \ref{sec:TPspec} and Sec. \ref{sec:gauss}, respectively.  This will be very effective with the upcoming multiple-frequency CMB surveys such as the Simons Observatory (SO) \cite{Ade_2019} and CMB-S4 \cite{abazajian2016cmbs4}. 

The simulation setup can use profile information from various hydrodynamical simulations or survey data to study the conversion phenomenon in different kinds of astrophysical systems, such as the circumgalactic medium, neutron stars, etc \cite{perna2012signatures,bondarenko2023neutron,lella2023protoneutron,yu2023searching}. Moreover, using the profile information, various astrophysical processes such as synchrotron emission, the SZ effect, Faraday Rotation, etc., can be simulated in this setup.

\section{Conclusion}
\label{sec:conclude}

The gravitational and astrophysical processes impact the hydrodynamics of the ionized plasma, which in turn affect the properties of a galaxy cluster, such as its magnetic field, electron density, temperature, etc. These processes lead to energy injection in the ICM, which leads to turbulence in these profiles \cite{subramanian2006evolving,schekochihin2006turbulence,xu2009turbulence,ensslin2006magnetic}. Thus, in a realistic scenario, the study of various cluster-based signals should take into account these disturbances, which affect the emissions from them.    

The strength, shape, and extent of the ALP distortion signal due to the photon-ALP conversion depends heavily on the magnetic field and electron density profiles in galaxy clusters. The ALP signal in temperature depends on the magnetic field and electron density magnitudes at the resonant locations ($m_a = m_{\gamma}$), but the polarization signal also depends on the transverse magnetic field direction along the line of sight at those locations.

In this work, we have used a simulation setup to realize galaxy clusters as a three-dimensional grid with separate grid lengths to account for the electron density variation scale and the magnetic field coherence length. For a galaxy cluster, we simulate the ALP distortion signal for ALP mass $10^{-13}$ eV, in both temperature and polarization for various cases of strength inhomogeneity and incoherence. We assign random magnetic field directions in different domains across all lines of sight in the cluster region on the sky. Also, we inject Gaussian inhomogeneities (which are a good approximation for relaxed clusters) in magnetic field and electron density, to check for its effects on the ALP signal \cite{subramanian2006evolving,schekochihin2006turbulence,xu2009turbulence,ensslin2006magnetic}. The cases of turbulence (standard deviation being the reported percentage of the mean) considered are the ones with no strength inhomogeneity (0\%), low inhomogeneity (10\%), and high inhomogeneity (30\%), on the cluster profiles. We have set the electron density grid size to $d_e = 100$ pc. If the coherence scale is not varied, it is set to $d_B = 100$ pc.

An increase in coherence length of the magnetic field from 100 pc to 1 kpc in orders of 10 leads to a suppression of power in polarization, but the temperature signal remains independent of it, as shown in Sec. \ref{sec:order}. An increase in magnetic field strength inhomogeneity affects both the temperature and polarization signals, as shown in Sec. \ref{sec:turb_B}, displaying a distinct increase in power at all multipoles for high inhomogeneity case. When together varied, as shown in Sec. \ref{sec:turb_Ball}, the combined effect is seen with the polarization signal getting much more affected due to the change in coherence length. The temperature signal changes only due to the magnetic field strength inhomogeneity. 

An increase in electron density turbulence changes not only the magnitude of the ALP signal but also its spatial extent and shape. The increase in resonances is accompanied by the probability of conversion getting weaker due to more fluctuating electron density, as explained in Sec. \ref{sec:turb_ne}. Finally, an increase in the strength inhomogeneity of both the magnetic field and electron density shows the combined effects of the two on the ALP signal as shown in Sec. \ref{sec:turb_neB}.

The observable signatures of these effects show up in the temperature and polarization 
distribution. Though CMB exhibits a Gaussian distribution, this signal can exhibit a non-Gaussian signal with a different spectral feature from any other signals. We show the correlation of the ALP temperature and polarization signals at different frequencies in Sec. \ref{sec:TPspec} and the deviation from the Gaussian distribution in Sec. \ref{sec:gauss} in both temperature and polarization. The non-Gaussian characteristic of the ALP signal and its evolution with inhomogeneity can be used to differentiate the ALP signal from the Gaussian CMB. This can be done by comparing the statistics of the region around galaxy clusters for the ALPs spectral distortion shape. In the absence of the ALP signal, the statistics will be Gaussian for almost all clusters. But if the ALP signal is present, then the ALP mass and coupling relation can be probed using the one-dimensional distribution of the signal depending on the angular size of the signal and its strength. 

The current understanding of the turbulence in electron density and magnetic field is primarily from simulations and some observations. As per hydrodynamical simulations, 5 to 30\% of the thermal energy in intracluster medium (ICM) is a result of turbulence motion of the ionized plasma, being specifically high for non-relaxed clusters \cite{dolag2005turbulent,brunetti2007compressible,schuecker2004probing,vazza2011massive,sunyaev2003detectability}. If unaccounted for, turbulence may lead to an order difference in the estimation of the ALP signal in temperature, and even more so in polarization. On the magnetic field side, the inferred coherence scale of the magnetic field from observations is about a few tens of kpc \cite{Carilli_2004, subramanian2006evolving, 2020ApJ...901..162H}. For such cases, the ALP distortion signal on the sky will be there in both temperature and polarization, and depolarization of the ALP distortion will be limited. These will show up in the temperature and polarization map, which can be captured using the auto-correlation and cross-correlation between these maps and also by inferring the induced non-Gaussian ALPs spectral distortion signal around galaxy clusters.

The simulation setup developed in this work for photon-ALPs signal will be integrated into the multi-band framework \texttt{SpectrAx} \cite{Mehta:2024sye}, which can use multi-frequency observations from radio, CMB, optical and X-ray surveys to infer the astrophysics of clusters, such as their magnetic field, electron density and temperature profiles, while also being able to constrain the photon-ALP coupling constant \cite{Mehta:2024sye}. This setup will enable a robust study of the ALP signal in temperature and polarization with inhomogeneous profiles, corresponding to more realistic astrophysical scenarios.

\begin{appendix}
\section{The case for high mass ALPs}
\label{sec:high mass}
\begin{figure}[h!]
     \centering
\includegraphics[height=6cm,width=12cm]{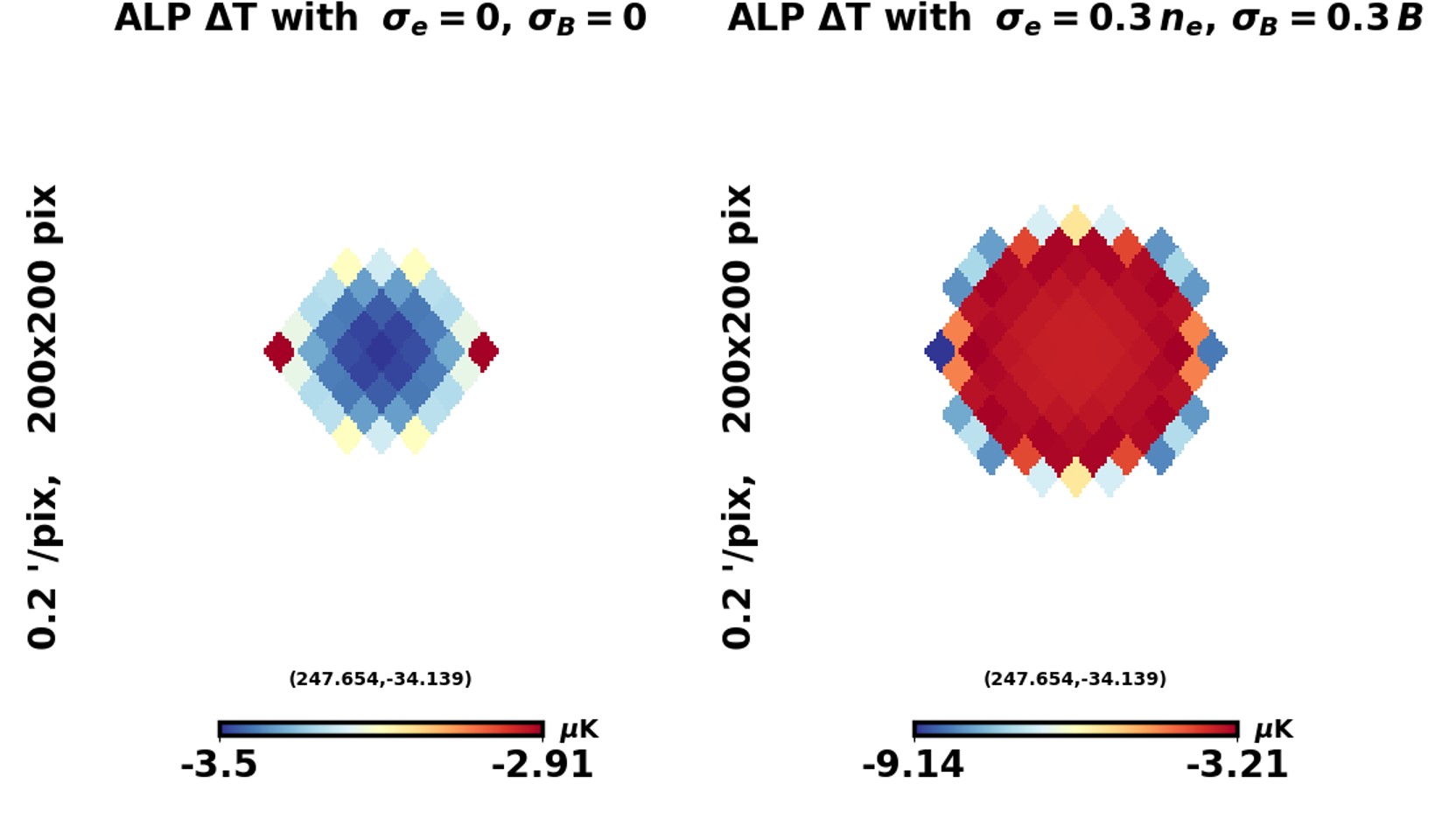}
    \caption{Variation in ALP temperature signal with change in both the electron density and magnetic field strength inhomogeneity for ALPs of high masses ($10^{-12} - 10^{-11}$ eV). Here domain size is $d_B = 1$ pc.}
       \label{fig:mhigh}
\end{figure}

\begin{figure}[h!]
     \centering
\includegraphics[height=7cm,width=11cm]{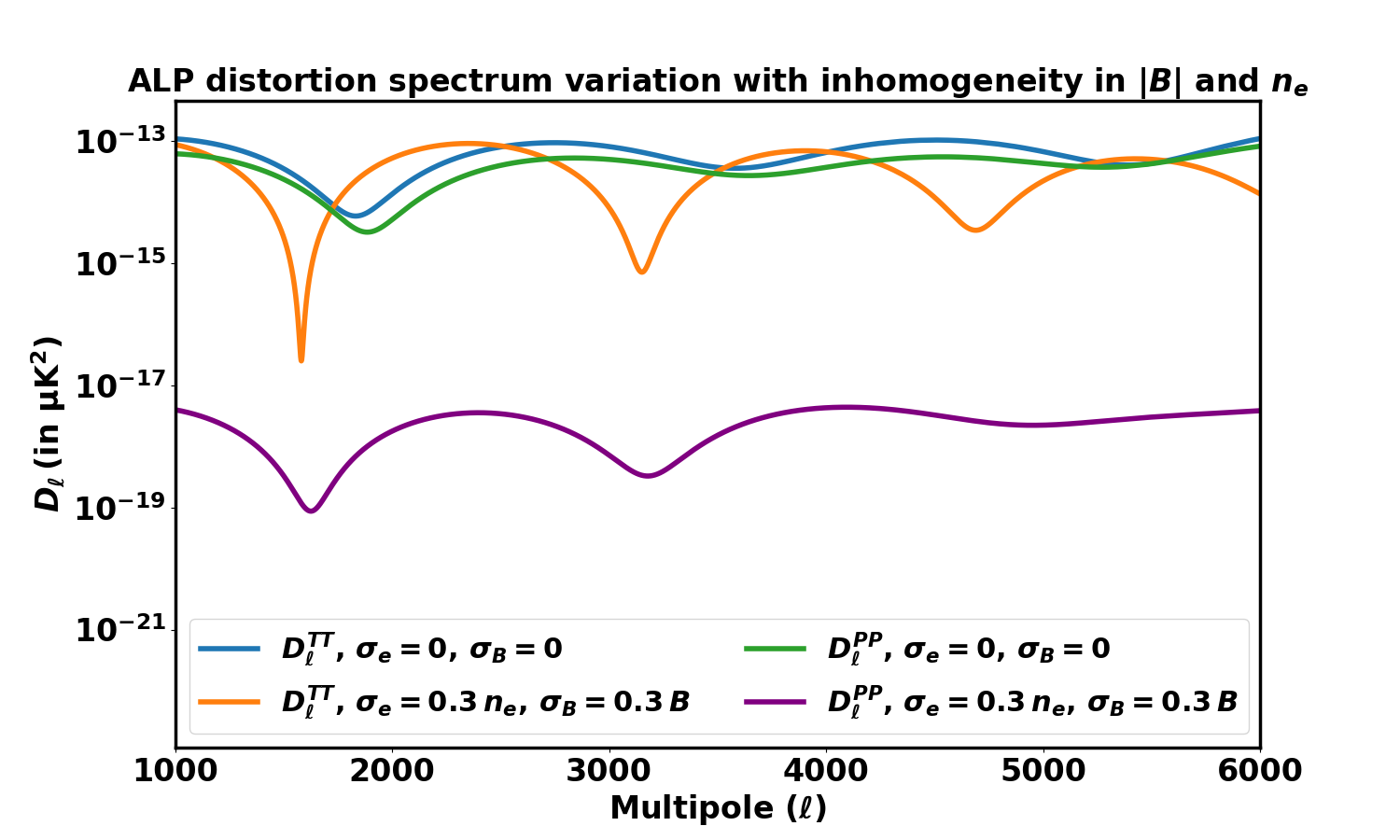}
    \caption{Variation in ALP TT and PP power spectra with change in both the electron density and magnetic field strength inhomogeneity for ALPs of high mass ($10^{-12}$ eV). Here domain size is $d_B = 1$ pc.}
       \label{fig:mhigh_TT}
\end{figure}

\begin{figure}[h!]
     \centering
\includegraphics[height=7cm,width=11cm]{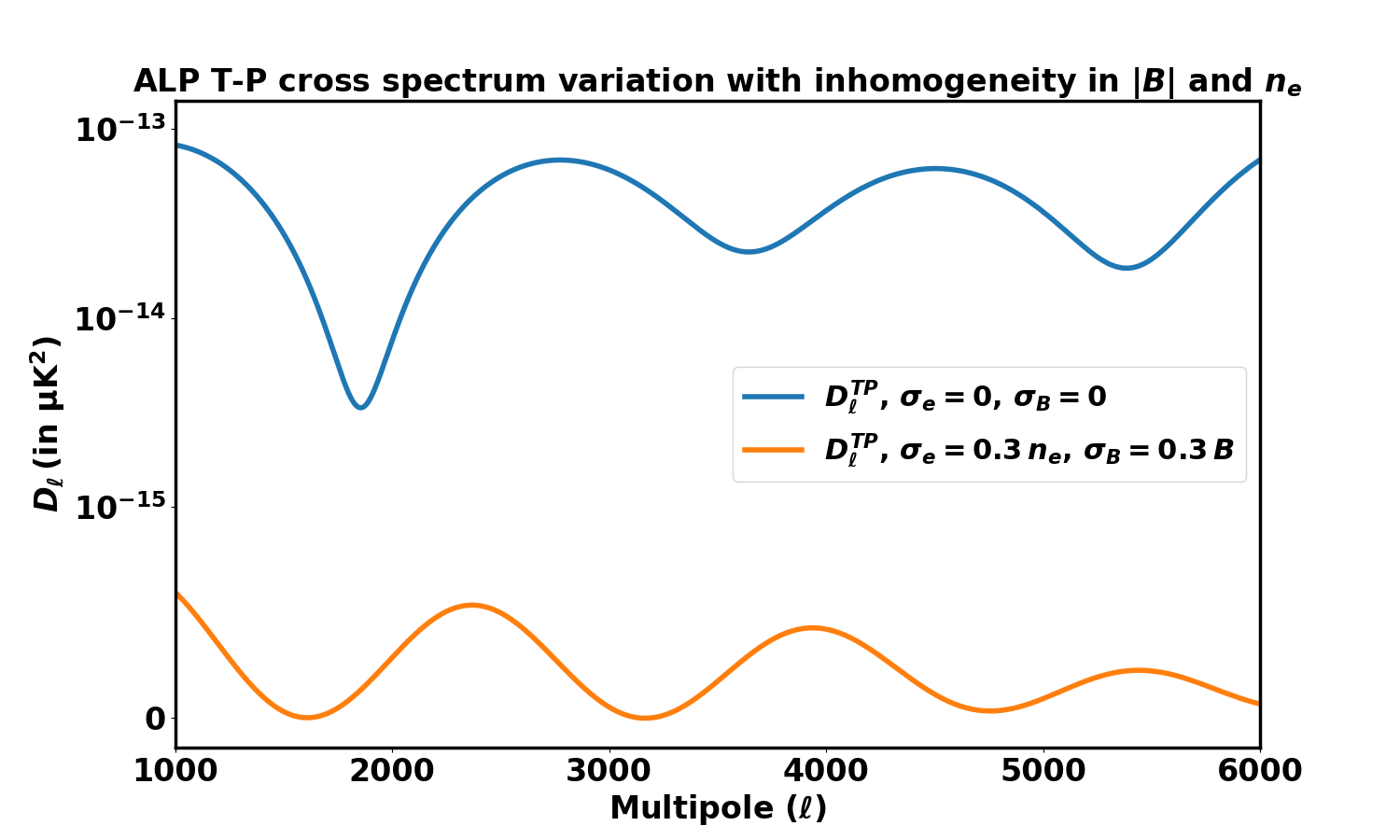}
    \caption{Variation in ALP TP cross spectrum with change in both the electron density and magnetic field strength inhomogeneity for ALPs of high mass ($10^{-12}$ eV). Here domain size is $d_B = 1$ pc.}
       \label{fig:mhigh_TP}
\end{figure}

\begin{figure}[h!]
     \centering
\includegraphics[height=7cm,width=11cm]{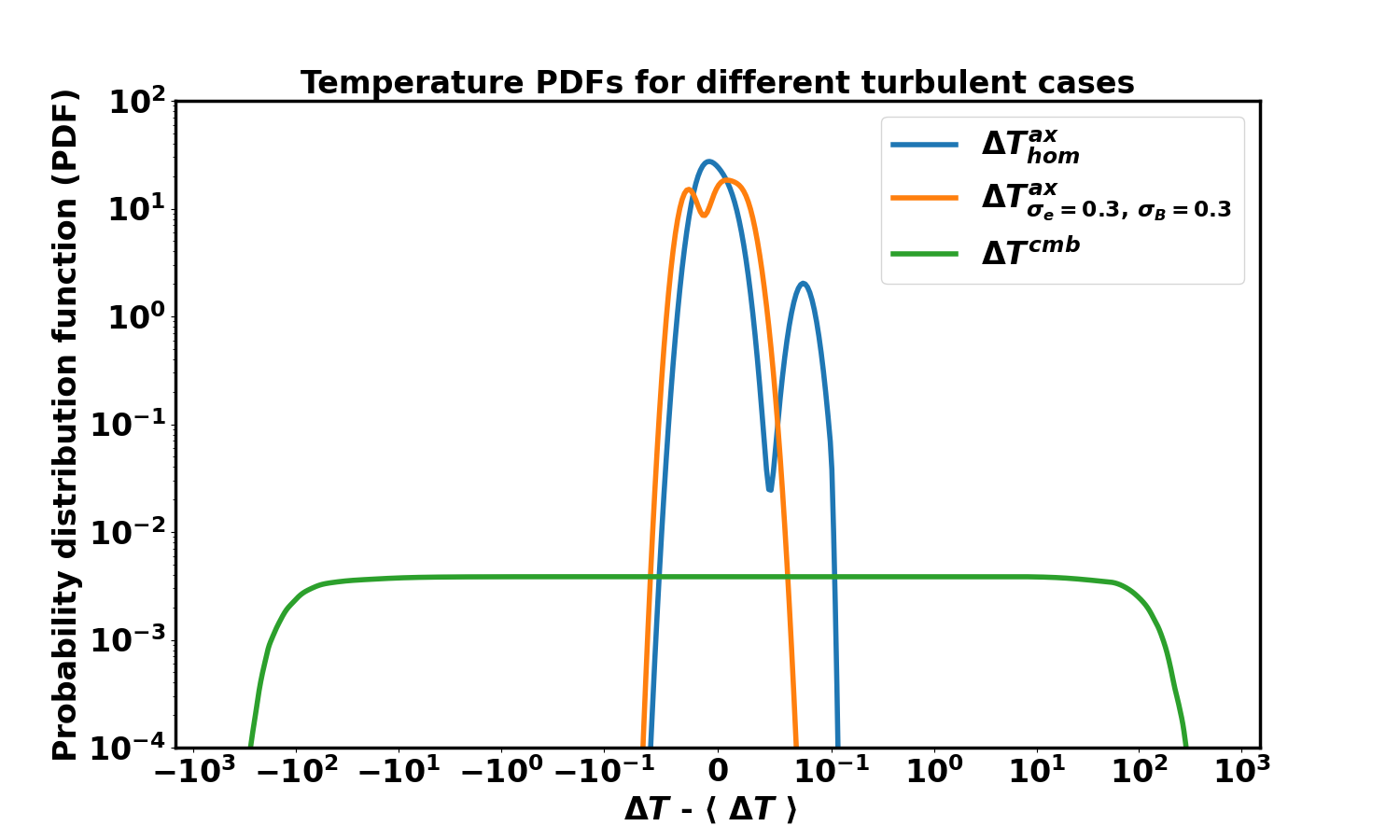}
    \caption{Temperature probability distribution functions (PDFs) of homogeneous and inhomogeneous cases for high mass ALPs ($m_a = 10^{-12}$ eV). Here $g_{a\gamma} = 10^{-11} \, \mathrm{GeV^{-1}}$.}
       \label{fig:T_high_pdf}
\end{figure}

We present a case for high-mass ALPs to show the working of the realization grid for a high-resolution scenario. The ALPs of high masses are formed in the inner regions of galaxy clusters, corresponding to higher electron densities, which are the regions where the turbulence is expected to be high \cite{Mukherjee_2020,Mehta:2024sye,Mehta:2024wfo}. The simulation setup can be scaled to study a part of the cluster where we expect the signal to be generated. We simulate the ALP signal for this case with a box length of 3 MPc and consider ALPs of mass $10^{-12}$ eV to be forming at the resonant locations. Also, we use a much-improved line of sight resolution of $d_e = d_B = 1$ pc, which will enable us to study the effects of turbulence at very small scales. 

We present the cases of no inhomogeneity and high inhomogeneity (standard deviation being 30\% of the mean) in both electron density and magnetic field in Fig. \ref{fig:mhigh} for the ALP temperature signal. The number of pixels where the signal is forming is quite low, but that can be improved by using a higher beam resolution. As for the earlier case, the extent of the signal disk increases due to turbulence in electron density. The mean signal in temperature decreases, but the fluctuations increase. This leads to an increase in power at some multipoles, while a decrease in others, as shown in Fig. \ref{fig:mhigh_TT}. The polarization power for the inhomogeneous case decreases drastically though, as a result of weak resonances, accompanied with depolarization of the signal. There is a drastic decrease in polarization power with inhomogeneity, which shows up also in the TP cross spectrum in Fig. \ref{fig:mhigh_TP}.  
The PDF of the temperature signal for the case of high-mass ALPs is shown in Fig. \ref{fig:T_high_pdf}, where the non-Gaussianity of the ALP signal changes with an increase in inhomogeneity, with a distinct reduction in the distribution at high values of the signal. 

Such a scaling can also be used to study the conversion signal in astrophysical systems of much smaller size as compared to clusters, such as neutron stars, galactic halos, etc \cite{perna2012signatures,bondarenko2023neutron,lella2023protoneutron,yu2023searching}. Thus, ALPs of a wide range of masses can be probed using this setup as a realization of an astrophysical system. The effect can be studied even better by using a lower beam size, with more pixels along the cluster line of sight, although that would lead to increased computational cost.  
\end{appendix}

\acknowledgments
    This work is a part of the $\langle \texttt{data|theory}\rangle$ \texttt{Universe-Lab}, supported by the TIFR  and the Department of Atomic Energy, Government of India. The authors express their gratitude to the TIFR CCHPC facility for meeting the computational needs. 
 Also, the following packages were used for this work: Astropy \cite{astropy:2013,astropy:2022,astropy:2018}, CAMB \cite{2011ascl.soft02026L}, NumPy \cite{harris2020array}, SciPy \cite{2020SciPy-NMeth}, SymPy \cite{10.7717/peerj-cs.103}, Matplotlib \cite{Hunter:2007} and HEALPix (Hierarchical Equal Area isoLatitude Pixelation of a sphere)\footnote{Link to the HEALPix website http://healpix.sf.net}\cite{2005ApJ...622..759G,Zonca2019}.

\bibliography{references.bib}
\end{document}